# Dynamics of Fluid Driven Autonomous Materials: Interconnected Fluid Filled Cavities to Realize Autonomous Materials

Yoav Matia[1†] & Amir D. Gat[2]
[1]Faculty of Mechanical Engineering, Cornell University Ithaca, New York 14853
[2]Faculty of Mechanical Engineering, Technion - Israel Institute of Technology
Technion City, Haifa, Israel 3200003

**Abstract**

The study of elastic structures embedded with fluid-filled cavities received considerable attention in fields such as autonomous materials, sensors, actuators, and smart systems. This work studies an elastic beam embedded with a set of fluid-filled bladders, similar to a honeycomb structure, which are interconnected via an array of slender tubes. The configuration of the connecting tubes is arbitrary, and each tube may connect any two bladders. Beam deformation both creates, and is induced by, the internal viscous flow- and pressure-fields which deform the bladders and thus the surrounding solid. Applying concepts from poroelasticity, and leveraging Cosserat beam large-deformation models, we obtain a set of three coupled equations relating the fluidic pressure within the bladders to the large transverse and longitudinal displacements of the beam. We show that by changing the viscous resistance of the connecting tubes we are able to modify the amplitude of oscillatory deformation modes created due to external excitations on the structure. In addition, rearranging tube configuration in a given bladder system is shown to add an additional degree of control, and generate varying mode shapes for the same external excitation. The presented modified Cosserat model is applied to analyze a previously suggested energy harvester configuration and estimate the efficiency of such a device. The results of this work are validated by a transient three-dimensional numerical study of the full fluid-structure-interaction problem. The presented model allows for the analysis and design of soft smart-metamaterials with unique mechanical properties.

†E-mail: ym279@cornell.edu

# 1. Introduction

This work examines how connections between different fluid-filled cavities in a solid structure can be used to control the response of the structure to external oscillatory excitations. The analysis focuses on an elastic beam embedded with a set of a fluid-filled bladders, interconnected via slender tubes (illustration of the examined configuration is presented in figure 1). We use a large deformation Cosserat continuum model for the solid, coupled with low-Reynolds-number flow in the slender tubes. Flow is induced by changes of the bladder volume, which are created by external forces or fluidic pressure. The fluid pressure within the bladders induces stress on the fluid-solid interface, creating local moments and normal forces, thus deforming the beam. This two-way coupling governs the dynamic response of such structures to external forces or forced fluidic pressurization.

Fluid structure interaction is an interdisciplinary research field focusing on the exchange of momentum, energy and chemical interactions between fluid and solid mediums. This exchange is governed by the conditions at the solid-fluid interface. Viscous flow in context of fluid structure interaction with an elastic medium has been studied as the driving mechanism at the heart of many physical systems [1]. In physiological flows, models of viscous flows in deformable slender elastic tubes provide insight into blood flow in arteries [2-6]. In another example, investigation of propulsion of a passive elastic flagella [7, 8] via a Stokeslet distribution, reveals the essential anisotropy in the velocity field at the center of drag-based swimming employed by microorganisms.

In geophysical flows, fluid structure interactions are relevant to a wide range of problems. Gravity flows through a deformable porous medium under a flexible surface [9], are a generalized configuration in which injected fluid spread under gravity, the matrix deforms and an upper elastic layer lifts up. Such models are relevant to seepage of contaminated plumes, or the displacement of one fluid by another in oil reservoirs to enhance recovery. Another subject is flow of ice driven by wind stress, impeded by water drag and shear stress relative to channel walls, which governs the dynamics and formation of wind-driven ice bridges [10]. Similarly, the dynamics of hydraulic fracturing backflows [11] can be modeled by using a simple representation of a fracture filled with a viscous fluid, confined in a bounded elastic medium. Other relevant research areas include dynamics of elastocapillary coalescence [12, 13], wrinkling instabilities in Hele-Shaw cells [14, 15, 16], viscous peeling [17, 18], Viscous-elastic interaction driven by non-uniform fluidic properties [19, 20], flow self-regulation via smooth-active control or elastic bi-stability [21, 22, 23], and impact mitigation driven by viscous-elastic dynamics [24]; where dominant balance between viscous forces and elastic bending and/or tension is shown to dominate phenomena propagation.

Of relevance to the current work are the fields of research of compliant vessels, poroelastic structures and soft robotics. The first subject focuses on the interaction of internal viscous flows with the dynamics of the bounding elastic medium. Heil & Pedley [2] provided a framework for axisymmetric deformable cylindrical shells conveying fluids. They lay out in detail the physical mechanisms involved in the interaction and examine it in three key steps. First, they illustrate tube deformation characteristics due to pressure void of through flow, then the effects of pressure distribution due to viscous flow are clearly apparent; last effects of the wall shear stress are included, providing insight on the effect of flow within such compliant vessels. Canic and Mikelic [5] studied the time-dependent flow governed by a given pressure drop between the inlet and the outlet, and provide a reduced set of visco-elastic equations for the effective pressure and the effective displacements.

The second subject focuses on the research of poroelastic structures where solid stress induces viscous flow which removes fluid from high-stress regions. Among many works in this field, Biot [25] prescribed the constitutive laws for a transversely isotropic poroelastic material. Cederbaum *et al.* [26] then expand on this, focusing on the unique features of poroelastic beam-shaped structures whose material is predominantly axially diffusive, and showed the mechanical behaviors of such structures to be greatly dependent on the diffusion patterns, which in turn dependence on pore anisotropy.

The third, and most related subject, is the rapidly expanding introduction of autonomous materials in soft robotics. Specifically, soft robots based on fluidic actuation are comprised of a compliant solid matrix laid with internal fluid-filled cavities, designed to undergo a set deformation pattern upon pressurization. In 1991 Suzumori et al. [27 was the first to present the field with a coherent theoretical framework in the form of a kinematic Denavit–Hartenberg type model with a dynamic transfer function for an infinite degree of freedom continuous soft actuator; introducing a pneumatic fiber reinforced rubber microactuators mounted on rigid bases. However, it is Ilievski et al. [28] and Shepherd *et al.* [29] that proposed a unique new class of robotic apparatuses. Unlike their predecessors, these were made entirely of an elastomeric solid matrix, emphasizing the advantages of soft robotics; simple elastomeric structures of appropriate design that can provide the function of a joint without the complexity of a multi-component mechanical structure. Many other researchers soon followed [30-52 , and while the use of fluids became the standard method of actuation, up to 2015, no models for fluid dynamics have been suggested in this field.

In our previous works [53, 54], , we presented a model for the effect of fluid mechanics on the actuation of soft robots for the first time. Examining an elastic beam embedded with a pressurized fluid-filled parallel-channel network, we derived the governing fluid-solid equations allowing leveraging viscosity to extend the capabilities of beam-shaped actuators. In this work, we generalize on previous configurations, modeling for any elastic beam-like appendage with an arbitrary set of fluid-filled interconnected cavities (illustration of the examined configuration is presented in Fig. 1). We present for the first time a two-way coupled fluid-structure interaction model for large transverse and longitudinal displacements, enabling new degrees of control relevant for soft robotics apparatuses and cellular metamaterials.

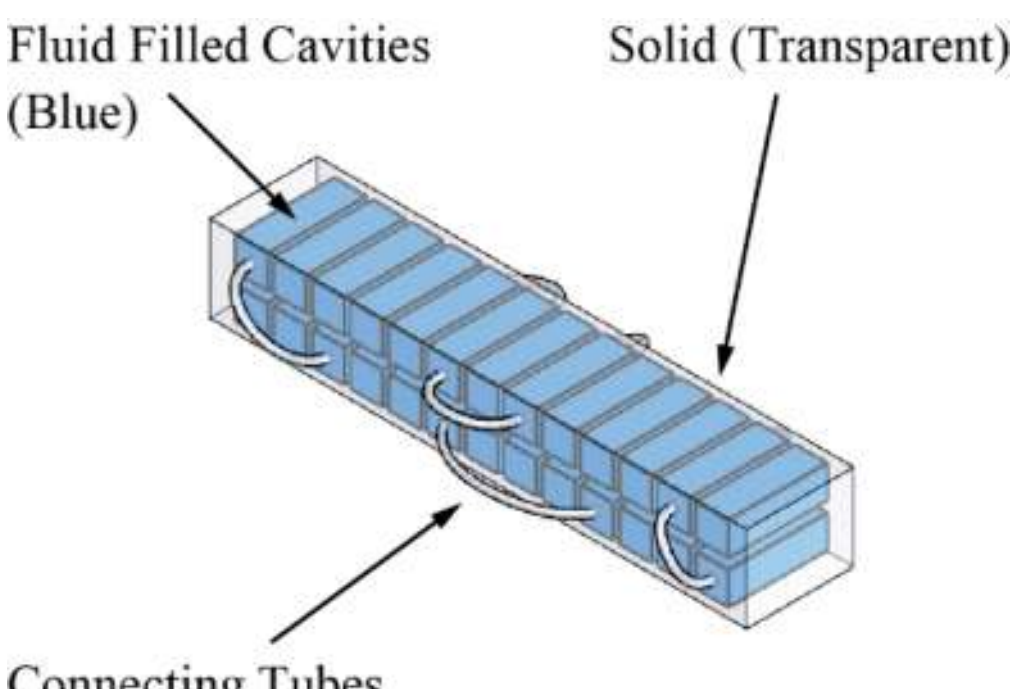

*FIG. 1.* ***Illustration of a slender beam with embedded interconnected fluid-filled bladders.***

# 2. Problem Formulation

In this work we present the effect of various configurations of fluidic connections between the fluid-filled bladders on the deformation modes of the beam due to external oscillatory excitations. A summary of used nomenclature is presented in Appendix A. We define vector variables using bold letters, direction vectors by hat notation, non-dimensional variables by tilde or capital letters, characteristic values by asterisk superscript and $s$ or $f$ subscripts for solid or fluid properties respectively. We define a lab frame of reference $(\hat{\boldsymbol{x}}_s, \hat{\boldsymbol{y}}_s, \hat{\boldsymbol{z}}_s)$ for the solid domain with a curvilinear length coordinate $\theta$ along the beam reference curve, and assign a position vector $\boldsymbol{x} = (x(\theta,t), z(\theta,t))$ pointing to the material point along the reference curve (neutral axis). Beam deflection is in the $\hat{\boldsymbol{x}}_s$ direction is $x(\theta,t) = x_0(\theta) + u_1(\theta,t)$ and extension in the $\hat{\mathbf{z}}_s$ direction is $z(\theta,t)=z_0(\theta)+u_3(\theta,t)$. For a beam which is straight at the unstrained state $x(\theta,t)=Const+u_1(\theta,t)$ and $z(\theta,t)=\theta+u_3(\theta,t)$. We define a fluidic field bladder-tube curvilinear frame $(\hat{\mathbf{x}}_f,\hat{\mathbf{y}}_f,\hat{\mathbf{z}}_f)$ defined such that $\hat{\mathbf{x}}_f$ is the streamwise direction, perpendicular to the $\hat{\mathbf{y}}_f-\hat{\mathbf{z}}_f$ plane. We define beam length $l_s$, beam height $h_s$, beam width $w_s$, Young's modulus $E$, solid density $\rho_s$, and parameters $f_e, f_i$ as solid field correction coefficients for cross sectional extension and bending stiffness, comparing the honeycomb structure to a full rectangular cross section structure with identical dimensions. Frames of reference and geometric dimensions are illustrated in Fig. 2.

Beam section-internal forces and moment resultants, due to traction, are normal force $N_e$ and moment $M_e$, and beam characteristic elastic-inertial time scale is $t_s^*=\sqrt{ml_s^4/EIf_i}[sec]$. Bladder location is defined by the integer function $R(x_f)$ where $R=1$ to indicate upper bladder and $R=-1$ for lower bladder. In addition, bladders are given an index $j$ by order of geometric position along $\theta$ such that $j\in[1,n/2]$ from left to right per row.

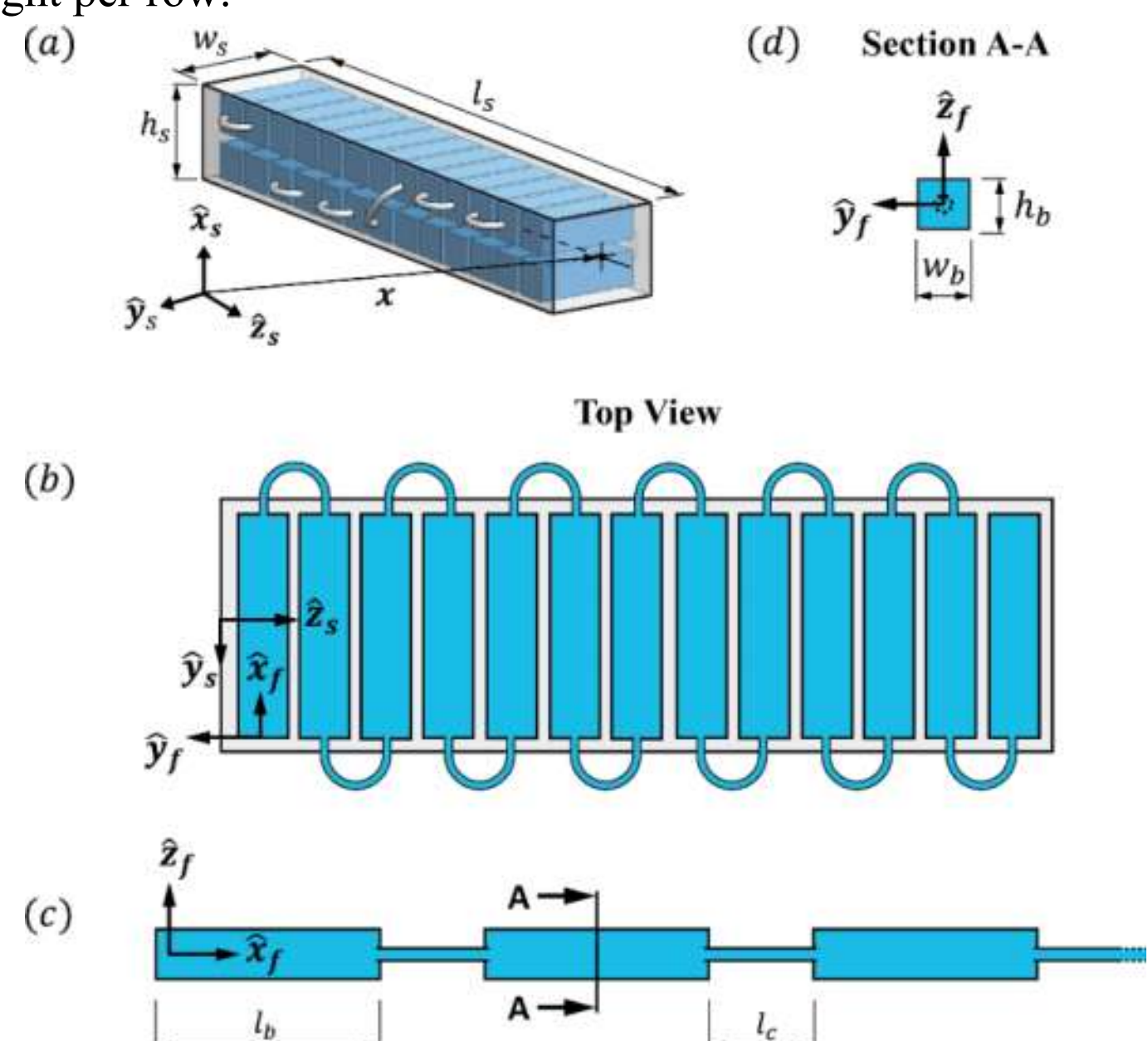

*FIG. 2.* ***Illustration of a solid-liquid composite beam structure with interconnected bladder array.*** *(a) Elastic beam with various connections between bladders, lab frame* $(\hat{\mathbf{x}}_\mathrm{s},\hat{\mathbf{y}}_\mathrm{s},\hat{\mathbf{z}}_\mathrm{s})$ *and position vector* $\mathbf{x}$ *pointing to material point along the reference curve (neutral axis). (b) Top view of a serpentine configuration and fluidic curvilinear frame* $(\hat{\mathbf{x}}_\mathrm{f},\hat{\mathbf{y}}_\mathrm{f},\hat{\mathbf{z}}_\mathrm{f})$*. (c) Mapping of the bladder and tubes into a straight continuous channel. (d) Definition of bladder cross section at rest (A-A).*

Bladder length, height and width are assigned $l_b$, $h_b$ and $w_b$, respectively. The slenderness of the tubes is given by $\varepsilon_1=2r_c/l \ll 1$, where $r_c$ is the tube radius and $l=l_c\, n_c$ is the total length of connective tubing, which is defined as the product of $l_c$ the length of a single connective tube and $n_c$ the total number of connective tubes in a given configuration. The fluidic field Viscous-elastic time scale is $t_f^*=\mu\left(\partial a_{p_1}/\partial p\right)\Big|_{p=p_0}/a_0^*\varepsilon_1^2\,[sec]$, where $\mu$ is the fluid dynamic viscosity, $\partial a_{p_1}(p)/\partial p$ is the dimensional change in bladder cross section area due to the fluid pressure and $a_0^*=\pi r_c^2[m^2]$ is the characteristic cross-section area of the fluidic domain i.e. bladder-tube array, at zero gauge pressure.

We characterize our system's two-way-coupled fluid-structure interaction by the FSI coupling coefficients in three distinct groups. The first describes how the fluidic domain pressure influences the geometry of the solid domain; these are $\partial a_{p_1}/\partial p$ defined above, the dimensional change in beam length per cross section per normalized pressure sum $\partial\zeta/\partial(\bar{p}/E)$, and the change in beam slope per cross section per normalized pressure difference $\partial\psi/\partial(p'/E)$. The second details how deformation to the solid domain influence bladder geometry and thus induce pressure in the fluidic domain; these are the change in fluidic cross-section area due to solid domain moment resultant $\partial a_{M_1}/\partial M_e$, the change in fluidic cross-section area due to solid domain normal force resultant $\partial a_{N_1}/\partial N_e$. The third details how bladder and connective tubing configuration affects the axial flux and includes the dimensional bladder permeability $q_1^b$ and connective tubing permeability $q_1^c$. The derivation of the solid field coefficients $f_i, f_e$ and fluid-structure coupling coefficients $\partial a_{p_1}/\partial p$, $\partial\zeta/\partial(\bar{p}/E)$, $\partial\psi/\partial(p'/E)$, $\partial a_{M_1}/\partial M_e$, $\partial a_{N_1}/\partial N_e$, $q_1^b$ and $q_1^c$ are determined by an equivalent reduced geometric model, either experimental or 3D FEM, whose properties are described in detail in Appendix A §A.4.1 and §A.4.2.

Next we define the normalized variables and coordinates. These include normalized beam curvilinear coordinate $\Theta = \theta/l_s$, curvilinear deflection axis $X_s = x_s/l_s$ and deflection variable $U_1 = u_1/l_s$, curvilinear extensional axis $Z_s = z_s/l_s$ and extension variable $U_3 = u_3/l_s$, moment resultant $\widetilde{M}_e = M_e/(E w_s h_s^3 f_i/12 l_s)$, normal force resultant $\widetilde{N}_e = N_e/E h_s w_s f_e[N]$, bladder-tube spatial coordinates $(X_f, Y_f, Z_f) = (x_f/l,\ y_f/h_b,\ z_f/h_b)$, viscous-elastic time $T = t/t_f^*$, fluid field pressure $P = p/E$. The effective pressure for slope and extension are defined $P'(X_f) = \left(P_d(X_f) - P_u(X_f)\right)$ and $\bar{P}(X_f) = \left(P_d(X_f) + P_u(X_f)\right)$ respectively, where $P_u = p_u/E$ and $P_d = p_d/E$ with $p_u$ and $p_d$ being fluidic pressures at upper and lower bladders respectively. Volume flow rate per bladder-tube array cross section is $Q = q/(\pi r_c^2\, E\varepsilon_1^2 l/\mu)$. We denote the non-dimensional permeability of the bladder as $Q_1^b = q_1^b/\tilde{C}^b r_{eff}^4$ and of the connective tubes $Q_1^c = q_1^c/\tilde{C}^c r_{eff}^4$ respectively, where $r_{eff}$ and $\tilde{C}^i \sim 4\pi$ ( $i = c, b$ ) are respectively the effective scale and dimensionless constant related to the configuration of the flow-path, and $q_1^i$ is defined by the relation $q = -((1/\mu)\partial p/\partial x_f) q_1^i$.

# 3. Results

The derivation and verification of the model are presented in detail in appendix A. In §A.1, the problem formulation and nomenclature are presented. In §A.2 we proceed to a formulation of the physical mechanism by which fluid flow and structural deformation interact. In §A.2.1, we start from the mass and momentum conservation equations, using order-of-magnitude analysis and obtain the non-linear diffusion equation governing the fluid field with two source terms due to solid section-internal normal force $\widetilde{N}_e$, and moment $\widetilde{M}_e$. In §A.2.2, we derive the intrinsic kinematic variables for curvature and lateral strain such that they include fluid induced terms and use them to formulate the constitutive relations for section-internal normal force resultant $\widetilde{N}_e$, shear force $\widetilde{V}_e$ and bending moment $\widetilde{M}_e$. In §A.2.3 we establish the Serret-Frenet triad coordinates associated with the beam reference curve and in §A.2.4 we derive the two way coupled solid field governing equations using an intrinsic formulation of a Cosserat rod, following [55, 56]. In §A.2.5-A.2.8 we complete the derivation of the governing equations by formulating the coordinate mapping between the fluid and solid domains, and last in §A.3 we provide a concise overview of the aforementioned model analysis.

## 3.1. Effect of the viscous resistance of the connecting network

This section presents the effect of modifying the resistance to flow between the fluid-filled bladders on the response of the beam to external excitations. Changing the resistance can be achieved by changing the fluid's viscosity, or changing the tubes geometry by valves or other methods, see illustration in Fig. 3. The physical parameters of the examined configuration are $l_s = 0.1[m]$, $h_s = 0.0216[m]$, $w_s = 0.050[m]$, $l_c = 0.020[m]$, $r_c = 0.001[m]$, $l_b = 0.046[m]$, and $h_b = w_b = 0.0078[m]$. The fluid used is water where $\rho_f = 1000[Kg/m^3]$, $\mu = 8.68 \cdot 10^{-4}[Pa \cdot sec]$, and the honeycomb structure material properties are $E = 2[MPa]$, $\rho_s = 950[Kg/m^3]$ and contains 20 bladders. The solid field coefficients and FSI coupling coefficients used in our configuration and the process of their calculation is detailed in Appendix A at §A.4.1 and §A.4.2.

The examined configuration is a cantilever beam, clamped at $\Theta = 0$. Starting from rest, we introduce an oscillatory excitation at the beam root $U_3(0,T) = U_{3,in}\, sin(2\pi\mathcal{F}_u T)$ where $U_{3,in} = 0.33$, $\mathcal{F}_u = f_{dim}t_f^* \approx 0.007$ and the dimensional oscillation frequency is $f_{dim} = 18[Hz]$. Oscillatory excitation parameters were selected such that resulting inertial axial-compression is significantly lower than the structure's limit for buckling; yet adequate to generate significant pressure gradients within the fluid-filled bladders.

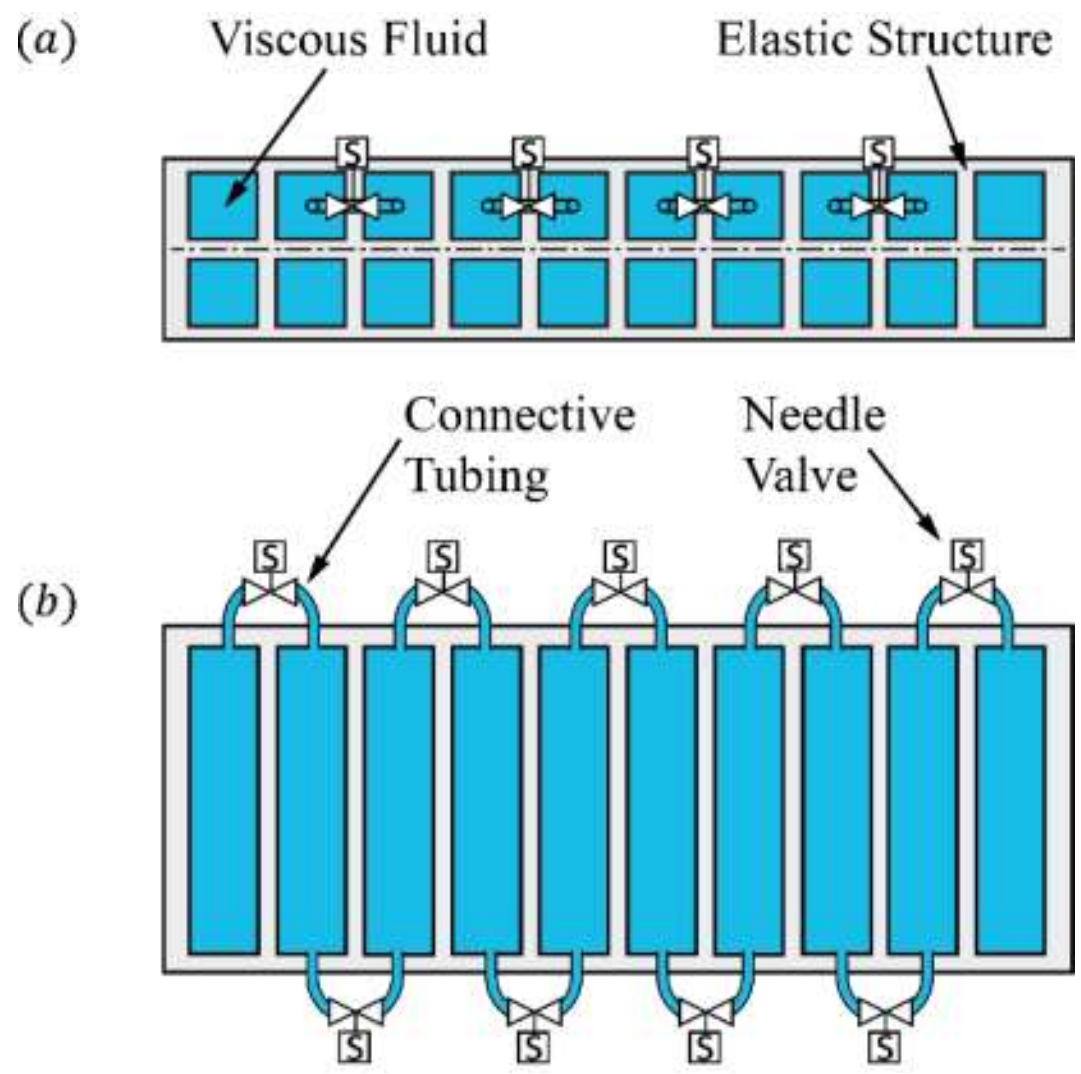

*FIG. 3.* ***Illustration of valves in a cantilever beam with embedded fluid-filled bladders.*** *(a) System front view. (b) System top view. Top row bladders are externally interconnected in a serpentine configuration. Each connecting tube has a needle valve, allowing to modify the viscous resistance of the tube. All lower bladders are unconnected.*

Fig. 4 presents the deformation of a beam with a serpentine tubing configuration for the top row bladders and no connections at the lower row of bladders, where the valves are used to set three different values of the permeability of the tubes connecting the bladders. In panel (a) we present an illustration of the configuration. Panels (bI-bIII) present deflection patterns for a full cycle. Panels (cI-cIII) present fluid pressure $P$ vs. the fluidic coordinate $X_f$. Panel (d) presents the correlation between bladders' normalized permeability $Q_1^b/Q_{1,open}^b$ and maximum induced deflection $U_1$. Column I presents results for unobstructed flow i.e. open valves, where $Q_{1,open}^b = 3.16 \cdot 10^{-3}$ and $Q_{1,open}^c = 1.37 \cdot 10^{-3}$. Column II presents partly restricted valves, represented by the reduced permeabilities of $Q^b_{1,open}/2$ and $Q^c_{1,open}/2$. Column III presents highly restricted flow with permeabilities reduced to $\sim Q^b_{1,open}/100$ and $\sim Q^c_{1,open}/100$.

For fully open valves (column III), the unobstructed tubes, with $Q^b_{1,open} = 3.16 \cdot 10^{-3}$ and $Q^c_{1,open} = 1.37 \cdot 10^{-3}$, enable for flow to equalize the pressure across the length of the top row. This occurs since viscous-elastic time is much smaller than the excitation time-scale. In contrast to the nearly uniform pressure at the upper row, the cavities at the lower row exhibit varying pressures defined from the localized values of $\widetilde{N}_e$ and $\widetilde{M}_e$ (see panel cIII). The pressure difference between lower and upper bladder rows induces a deformation mode perpendicular to the external excitation (i.e., oscillating the system in $U_3$, the pressure field excites the oscillations in $U_1$). Setting the valves to increase viscous resistance to $Q^b_{1,open}/2$ and $Q^c_{1,open}/2$ significantly reduces the oscillations in $U_1$ (column II). Further increasing the viscous resistance to $\sim Q^b_{1,open}/100$ and $\sim Q^c_{1,open}/100$ (column I), effectively inhibits flow between bladders. This eliminates the excited perpendicular mode in $U_1$ and the beam remains straight during the oscillation (Fig. 4bI). Fig 4d. presents the relation between the maximal deformation and the viscous resistance, via the ratio $Q^b_1/Q^b_{1,open}$. The transition between a region of negligible effect (column I) and maximal effect (column III) occurs in $Q^b_1/Q^b_{1,open}$ values between $\approx 0.3$ and $\approx 0.7$, representing the region in which viscous-elastic time-scale is similar to the time-scale of the external excitations. Thus, changing the network resistance allow to modify the amplitude of the induced deflections gradually between the limit of uniform pressure in all bladders and the limit of disconnected bladders with different pressures.

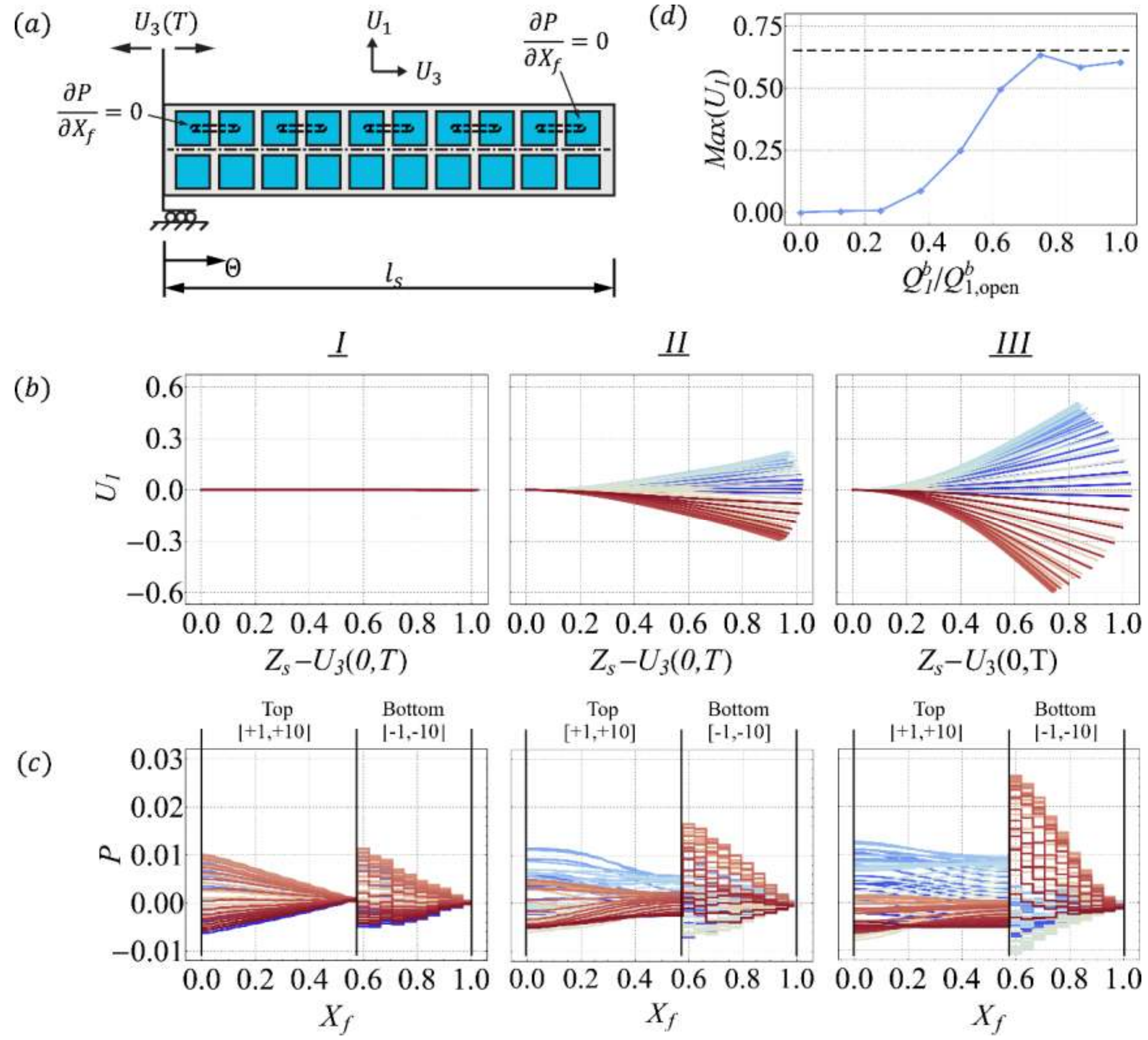

*FIG. 4.* ***Effect of the viscous resistance of the connecting network.*** *Cantilever solid-liquid beam clamped at* $\Theta = 0$*, with top row bladders interconnected in a serpentine configuration. Viscous-resistance is modified by valves or similar methods. The beam is excited by oscillatory excitation at the clamped end* $\mathrm{U_3(0, T) = U_{3,in} \sin(2\pi\mathcal{F}_u \cdot T)}$ *where* $\mathrm{U_{3,in} = 0.33}$ *,* $\mathcal{F}_\mathrm{u} = \mathrm{f_{dim} \cdot t_f^*} \approx 0.007$ *and the dimensional oscillation frequency is* $\mathrm{f_{dim} = 18[Hz]}$*. (a) System section cut view. (b) Beam deflection. (c) Fluid pressure P vs bladder-tube coordinate* $\mathrm{X_f}$*. (d) Scaled permeability* $\mathrm{Q_1^b/Q_{1,open}^b}$ *vs maximum induced deflection* $\mathrm{U_1}$*. Results for open valve i.e. unobstructed flow* $\mathrm{Q_1^b = 3.16 \cdot 10^{-3}}$ *and* $\mathrm{Q_1^c = 1.37 \cdot 10^{-3}}$ *(column III), partly restricted flow* $\mathrm{Q_1^b/2}$ *and* $\mathrm{Q_1^c/2}$ *(column II) and highly restricted flow* $\mathrm{\sim Q_1^b/100}$ *and* $\mathrm{\sim Q_1^c/100}$ *(column I) are presented. Bladders'* $(\mathrm{R \cdot j})$ *index along* $\mathrm{X_f}$ *coordinate is noted in square brackets in panel (c). Time evolution is presented via transition from blue to red.*

## 3.2. Effects of the wiring configuration

The previous section examined the effect of viscous resistance of the connecting tubes on the response of the beam to external excitations. Valves were suggested as a method to modify the viscous resistance. A similar approach may be used to change the geometry of the connecting network. By closing some valves and opening others, the wiring of the connecting network can be changed. In this section three different wiring configurations are examined, as illustrated in Fig. 5a. All include two disconnected networks, marked by blue and red lines (see appendix A §A.2.5 and in Fig. A.3b & A.3c). Configuration I includes an upper and lower serpentine connections, which are both closed at the inlet and outlet. Configuration II includes a similar setup, but with the lower channel inlet and outlet open with gauge pressure set to zero. In configuration III, the upper and lower channels are switched at the fourth bladder. The conditions at the ends are identical to configuration II. All solid and fluid properties are identical to section 3.1, where the excitation amplitude was reduced to $U_{3,in} = 0.12$ at dimensional frequency $f_{dim} = 20[Hz]$.

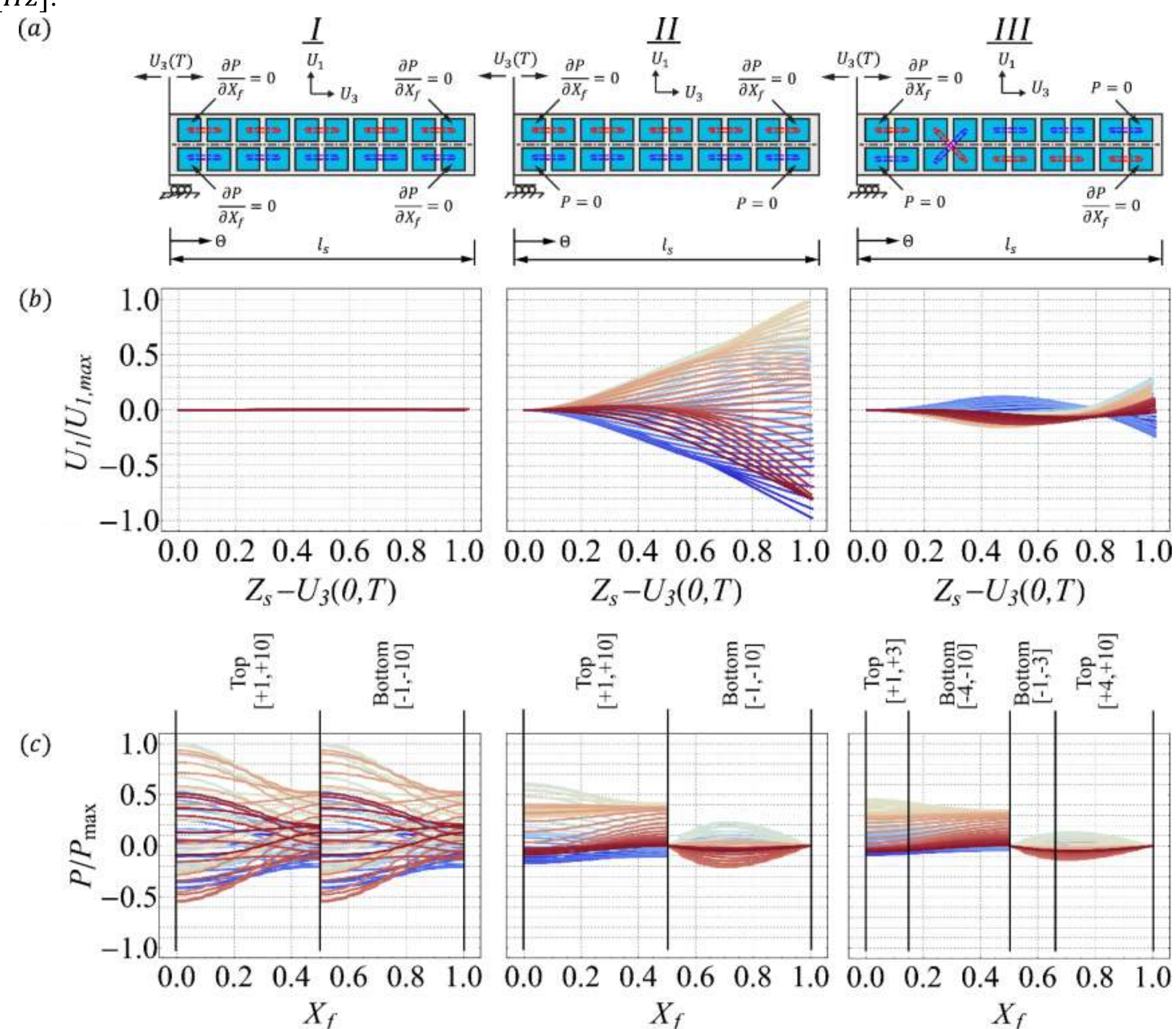

*FIG. 5.* ***Effects of the wiring configuration. (a) Network wiring illustrations.*** *(b)* $U_1/U_{1,max}$ *scaled deflections. (c) Scaled fluid pressure* $P/P_{max}$ *vs* $X_f$. *Column I presents an upper and lower serpentine connections, which are both closed at the inlet and outlet. In Column II the lower channel inlet and outlet are open with gauge pressure set to zero. Column III presents wiring where the upper and lower channels are switched at the fourth bladder. Bladder* $(R \cdot j)$ *index range along* $X_f$ *coordinate are noted in square brackets in panel (c). Time evolution is presented via transition from blue to red. The excitation at the cantilever root is* $U_3(0,T) = U_{3,in} \sin(2\pi\mathcal{F}_u \cdot T)$ *where* $F_u = t_f^* \cdot f_{dim} \approx 0.0078$, *with* $f_{dim} = 20[Hz]$ *and* $U_{3,in} = 0.12$. *The scaling parameters are* $P_{max} = 7.3 \cdot 10^{-3}$ *and* $U_{1,max} = 0.137 \cdot 10^{-3}$.

Panels (b) and (c) in Fig. 5 present the beam deflection and fluid pressure, respectively, for all three configurations. For the symmetric configuration I, the displacement $U_1$ is zero since fluid pressure in upper and lower bladders is identical. In configurations II and III, once boundary conditions vary between upper and lower arrays, pressure differences between the upper and lower bladders create significant deformation. In configuration II, setting the pressure to zero at the fluidic channel ends significantly reduces the magnitude of pressures within lower bladders. Thus, the greater pressures at the upper bladders determine beam deflection. Therefore, all of the beam experiences negative curvature for expansion in $U_3$ and positive curvature for compression in $U_3$. In configuration III, we keep the same boundary conditions but switch the networks between the upper and lower bladders. Thus, the beam curvature is inversed after the switch, and a deflection pattern resembling a second-mode is created, (see panel 4.5bIII). Hence, changing the network wiring allow to control not only the amplitude, as seen in Figs. 3 and 4, but also the deflection pattern due to external oscillations.

## 3.3. Energy Harvesting

Oscillation dynamics of elastic sheets due to external forces are studied extensively in the context of wind based energy harvesting. These oscillations are determined by interaction between the non-steady external aerodynamics and the elastic and inertial dynamics of the beam. Works in this field study flow dynamics, instabilities and flutter [57-59] and their relation to efficiency of wind energy harvesting [60-63]. Commonly, piezoelectric materials, embedded within the elastic sheet, are studied as a mechanism to harvest energy from the sheet elastic deformations [64, 65]. However, as suggested in [66, 67], fluidic-embedded beams can be an alternative to piezoelectric materials for such energy harvesting applications. This section utilizes the derived model to examine the effect of interconnections between different bladders on the efficiency of such a device.

An illustration of energy harvesting configurations with a fluid-filled beam is presented in Fig. 6a,b for two different networks connecting the bladders. The external excitations deform the beam, thus deforming the internal bladders and force fluid from (or into) the beam; thus, the geometry and properties of the connecting tube network can be used to change the sheet oscillations and optimize the efficiency of wind energy harvesting. Connecting the bladders at the beam base to a generator allows to harvest energy from all bladders connected to the network. To increase the system efficiency for a given external load, all bladders with reduced volume (at a specific time) should be connected to a single network. Similarly, all bladders with increased volume due to external forces should be connected to a second network. Thus, since different external loads yield different deformation patterns, different loads require different wiring to achieve improved efficiency.

We define the energy harvested by the fluid via

$$\mathcal{W}_{fluid} = \int_{\Delta T} Q \cdot \Delta P \, dT = \int_{\Delta T} \left( \frac{\partial P}{\partial X_f} Q_1^b \right) ( P_u - P_d) \, dT \,, \tag{3.1}$$

where fluid power output is given by $\mathcal{P}_{fluid} = Q \cdot \Delta P = \left( \frac{\partial P}{\partial X_f} Q_1^b \right) ( P_u - P_d)$, $\mathcal{p}_{fluid}^* \sim q^* p^* = 4E^2 \pi r_c^4 / \mu$, $\Delta T$ is the examined time period, $Q$ is flow rate, $P_u$, $P_d$ are the pressure at the upper and lower inlets respectively, and $\mathcal{W}_{fluid}$ is scaled by $\mathcal{w}_{fluid}^* \sim 4\pi r_c^4 E^2 t_f^* / \mu l$. The work done by the external forces is

$$\mathcal{W}_{Input} = \int_{\Delta T} \int_{\theta} B_{c_x} \frac{\partial U_1}{\partial T} \, d\theta \, dT \,, \tag{3.2}$$

where $\mathcal{W}_{Input}$ is scaled by $\mathcal{w}_{input}^* \sim b_x^* u_1^* l_s$. The connection to the turbine sets boundary conditions of $\partial P_u / \partial X_f = \tilde{\kappa}_{turb} (P_u - P_d)$ and $\partial P_d / \partial X_f = -\tilde{\kappa}_{turb} (P_u - P_d)$, where $\tilde{\kappa}_{turb}$ is the turbine non-dimensional conductance, scaled by $\kappa_{turb}^* \sim \mu q^* / q_1^{b*} p^*$. Thus we can define the direct cycle-averaged efficiency

$$\bar{\eta} = \frac{\mathscr{w}_{fluid}}{\mathscr{w}_{Input}}, \tag{3.3}$$

as well as the instantaneous efficiency

$$\eta(t) = \frac{\mathscr{p}_{fluid}}{\mathscr{w}_{Input}/\Delta t}, \tag{3.4}$$

where $\mathscr{w}_{Input}/\Delta t$ is the time-averaged power input.

In Fig. 6, we examine the effects of matching the wiring configuration to the external force distribution. We present a cantilever beam under an oscillating external load $B_{c_x} = 2.17 \cdot sin(2\pi\mathcal{F}_u T)\, cos(2\pi\lambda_{wave}\Theta)$, where $b_x^* = 46.189[N/m]$, $F_u = t_f^* \cdot f_{dim} \approx 3.93 \cdot 10^{-4}$, $f_{dim} = 1[Hz]$. Panel (a) and (b) illustrate wiring for parallel and crossover configuration respectively, with full-wave bridge rectifier and turbine. In panel (cI, cII) we present force distribution plot, where time evolution is presented via transition from blue to red. In panels (dI, dII) we plot the instantaneous energy harvesting efficiency $\eta(\tau)$ for $\lambda_{\text{wave}} = 0$ and $\lambda_{\text{wave}} = 0.95$ in columns *(I, II)* respectively for both wiring configurations. All solid and fluid properties are identical to §3.1 and $\tilde{\kappa}_{turb} = 0.1875$.

For a uniformly distributed load expressed by $\lambda_{\text{wave}} = 0$ (see panel 4.6c) positive external force $B_x$ causes all upper bladders to contract as the lower bladders expand, and vice versa for uniformly negative $B_x$. Thus, the optimal wiring will connect the upper and lower bladders in two separated networks, as illustrated in panel (a). Panel dI confirm the efficiency of such a wiring, specifically in contrast to the crossover wiring presented in panel (b). However, for $\lambda_{\text{wave}} = 0.95$ the external load changes the curvature of the beam, and the crossover wiring presents better efficiency in this case. Thus, by matching bladder wiring configuration to a given set of boundary conditions and external force distribution, energy harvesting efficiency can be maximized per external actuation, yielding efficiencies of up to 40%.

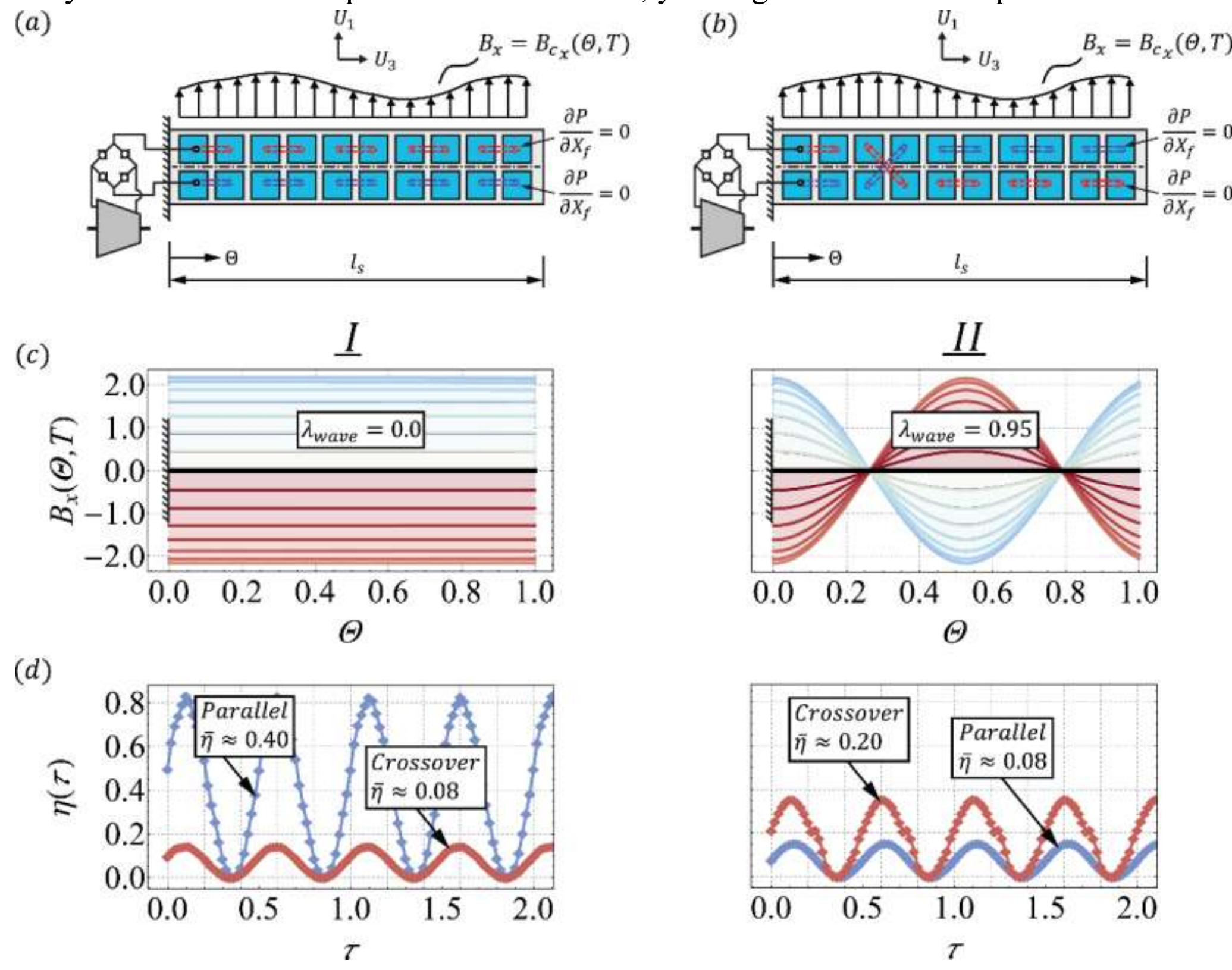

*FIG. 6.* ***Illustration of an energy harvesting system, setup and results.*** *Results are presented for oscillating force distribution* $\mathrm{B_{c_x}} = 2.17 \cdot \sin(2\pi\mathcal{F}_u \mathrm{T}) \cos(2\pi\lambda_{\mathrm{wave}}\Theta)$*, where* $\mathrm{b_x^*} = 46.189[\mathrm{N/m}]$*,* $\mathrm{F_u} = \mathrm{t_f^*} \cdot \mathrm{f_{dim}} \approx 3.93 \cdot 10^{-4}$*,* $\mathrm{f_{dim}} = 1[\mathrm{Hz}]$*. Panel (a) and (b) show parallel and crossover wiring configurations respectively, as well as full-wave bridge rectifier and turbine. (c) External force distribution plots. Time evolution is presented via transition from blue to red. (d) Instantaneous energy harvesting efficiency* $\eta(\tau)$ *for* $\lambda_{\mathrm{wave}} = 0$ *and* $\lambda_{\mathrm{wave}} = 0.95$ *in columns (I, II) respectively.*

# 4. Concluding Remarks

In this work, we present a modified Cosserat model of an elastic beam embedded with fluid-filled cavities that are interconnected by a network of tubes in an arbitrary configuration; allowing for the first time, to model for large deformation dynamics of an elastic beam containing fluid-filled cavities, while including full coupling between the solid and fluid domains. Conditions at the solid-fluid interface governs the time evolution behavior of a solid volume with internal fluidic cavities. Structure deformation both creates and is induced by internal viscous flow. A change in axial flux balances changes to cavities' cross-section; pressure gradients develop in the fluid in order to conserve mass, and induce stresses at the solid-fluid interface. These create local moments and normal forces, deforming the surrounding solid and vice versa; completing the cycle of the viscous-elastic transient dynamics of such systems.

The effect of the connecting network on the beam response to external excitations was examined. Modification of viscous resistance or geometry of the connecting tubes (via valves or other methods) was shown to significantly modify both the amplitude and shape of the induced deformation patterns. The presented approach thus allows to model and design structures with dynamically changing mechanical properties, via opening or closing of valves connecting the fluid-filled cavities. All the above is presented for the first time in the fields of autonomous materials and soft-smart systems. Configurations similar to the examined beam containing multiple fluid-filled cavities are also common to the field of fluidic-driven soft robotics. The presented model can be used as a basis for calculation of the external forces on the beam, via measurement of the fluidic pressure, thus using fluidic-driven soft actuators as sensors. Last we show an applicative illustration, for the optimization of wind energy harvesting by natural oscillations of elastic sheets. These oscillations are determined by the interaction between the non-steady external aerodynamics and the elastic and inertial dynamics of the beam. The wind-based deformation of the elastic sheet pressurizes the embedded fluid, which both allows to directly to harvest energy and modify the sheet oscillations. Thus, the geometry and properties of the network of connecting tubes can be used to change the sheet oscillations and optimize the efficiency of wind energy harvesting.

The aim of this research is to introduce a comprehensive fluid dynamics model to the fields of autonomous and smart materials; enabling engineers and scientists to leverage fluidic phenomena in respective fields. The solid-fluid dynamics' of such systems intrinsically possess two dominant time scales: a viscous-elastic time scale the fluid field $t_f^*$, characterizing the diffusion of pressure along the fluid field, and the inertial-elastic time scale, $f_s^*$, characterizing solid field transient dynamics. Three dynamic regimes raise from these two time scales. The first limit is where $t_s^*/t_f^* \gg 1$, the viscous-elastic phenomena propagate significantly faster than solid inertial response, resulting in a uniform pressure field where viscous effects are negligible, and system dynamics is dominated by solid inertia. Second where $t_s^*/t_f^* \ll 1$; in this opposite limit, solid inertia is negligible, and deformation reflects the viscous-elastic dynamics. Last and most complex is $t_s^*/t_f^* = O(1)$; where both fluid viscosity and solid inertia govern structure dynamics. We focus on the last regime where $t_s^*/t_f^* = O(1)$.

The generalized mathematical model presented can be readily reduced to the two other dynamic regimes by applying the limiting assumptions of $t_s^*/t_f^* \gg 1$ or $t_s^*/t_f^* \ll 1$, thus eliminating respective terms. The generalized form presented in this work, provide for the first time a comprehensive fluid-structure interaction model suited to any of the three dynamic regimes aforementioned, as well as for a wide and diverse range of beam-like soft structures common to the field of smart materials and soft robotics owing a unique set of controllable mechanical properties, and providing insight into the previously unexplored leveraging of actuating fluid viscosity in respective fields.

# 5. Acknowledgments

This research is funded by ISRAEL MINISTRY OF SCIENCE, TECHNOLOGY AND SPACE and the ISRAEL SCIENCE FOUNDATION (Grant No. 818/13).

# 6. Author Contributions

Y. Matia and A.D. Gat jointly conceived the study, created the analytic model, developed analytical tools, analyzed data, performed numerical calculations, and wrote the paper; A.D. Gat supervised the project and manuscript.

# 7. Competing Financial Interests

The authors declare no competing financial interests.

# Appendix A

# A. Supplementary Materials

In this section we provide a detailed step by step formulation of the analytic model whose results are presented in §3.

## Nomenclature

We define vector variables by bold letters, direction vectors by hat notation, non-dimensional variables by tilde or capital letters and characteristic values by asterisk superscripts.

***System physical parameters***

|  |  |
|---|---|
| $l_s$ | Beam length. |
| $h_s$ | Beam height. |
| $w_s$ | Beam width. |
| $E$ | Beam material modulus of elasticity. |
| $\nu$ | Beam material Poisson ratio. |
| $\rho_s$ | Beam material mass density. |
| $f_m$ | Solid mass fraction. |
| $f_e$ | Cross sectional extensional stiffness correction coefficient, comparing the honeycomb structure to a full rectangular cross section structure with identical dimensions. |
| $f_i$ | Cross sectional flexural stiffness correction coefficient, comparing the honeycomb structure to a full rectangular cross section structure with identical dimensions. |
| $m$ | Beam mass per unit length. |
| $I$ | Beam cross section moment of inertia. |
| $y^{11}, Y^{11}$ | Gyration radius squared, and its non-dimensional form. |
| $\psi$ | Change in beam slope due to a single pressurized bladder (intrinsically non-dimensional). |
| $\zeta, \tilde{\zeta}$ | Change in beam length due to a single pressurized bladder, and its non-dimensional form. |
| $\phi, \Phi$ | Structure length-wise bladder density, and its non-dimensional form. |
| $l_b$ | Length of a single bladder segment. |
| $h_b$ | Height of a fluidic bladder cross section. |
| $w_b$ | Width of a fluidic bladder cross section. |
| $n$ | Total number of bladder (fluidic cavities) in the honeycomb structure. |
| $\rho_f$ | Fluid domain material mass density. |
| $\mu$ | Fluid dynamic viscosity. |
| $\varepsilon_1$ | Small parameter representing slenderness of the fluidic domain. |
| $r_c$ | Connective tubing radius. |
| $l$ | Total length of connective tubing configuration. |
| $l_c$ | Length of a single connective tube. |
| $n_c$ | Total number of connective tubes in a given configuration. |
| $\sigma_p$ | Ratio of $a_0^*$ the characteristic cross-section at gage pressure to $a_1^*$ the characteristic change in bladder cross-section. |

| $r_{eff}$ | Effective dimensional scale related to the configuration of the flow-path i.e. averaged hydraulic radii of bladder and connective tubing. |
|---|---|
| $\tilde{C}^b$ | Dimensionless constant, related to the bladder flow-path i.e. shape of the bladder cross-section. |
| $\tilde{C}^c$ | Dimensionless constant, related to the connective tubing flow-path i.e. shape of connective tube cross-section. |

***Variables and arguments***

| $t, T, T_s$ | Time dimension, viscous-elastic non-dimensional time, and Inertial-elastic non-dimensional time. |
|---|---|
| $\theta, \Theta$ | Curvilinear length coordinate along the beam reference curve, and its non-dimensional form. |
| $(\hat{\boldsymbol{x}}_s, \hat{\boldsymbol{y}}_s, \hat{\boldsymbol{z}}_s)$ | Solid domain lab frame of reference. |
| $(x_s, y_s, z_s)$ , $(X_s, Y_s, Z_s)$ | Solid domain lab frame coordinates, and their non-dimensional form. |
| $\boldsymbol{x}$ | Lab frame position vector for the beam reference curve (i.e. the neutral axis). |
| $({}^s\boldsymbol{e}_n, {}^s\boldsymbol{e}_\tau, {}^s\boldsymbol{e}_t)$ , $({}^S\boldsymbol{E}_n, {}^S\boldsymbol{E}_\tau, {}^S\boldsymbol{E}_t)$ | Spatial Serret-Ferent triad (at time $t$) associated with the current position along the reference curve in a curvilinear frame using the unit tangent ${}^s\boldsymbol{e}_t$ pointing the direction of motion, the unit normal ${}^s\boldsymbol{e}_n$ , unit binormal ${}^s\boldsymbol{e}_\tau$, and their non-dimensional form. |
| $(\boldsymbol{d_1}, \boldsymbol{d_2}, \boldsymbol{d_3})$ | Spatial director (strain vector at time $t$), in the curvilinear frame of reference $({}^s\boldsymbol{e}_n, {}^s\boldsymbol{e}_\tau, {}^s\boldsymbol{e}_t)$. |
| $(\boldsymbol{D_1}, \boldsymbol{D_2}, \boldsymbol{D_3})$ | Material director (strain vector at time $t = 0$), in the curvilinear frame of reference $({}^s\boldsymbol{e}_n, {}^s\boldsymbol{e}_\tau, {}^s\boldsymbol{e}_t)$. |
| $u_1, U_1$ | Beam deflection displacement and its non-dimensional form. |
| $u_3, U_3$ | Beam extension displacement and its non-dimensional form. |
| $\lambda_s, \tilde{\lambda}_s$ | Total structure measure of stretch, Intrinsic kinematic variable, and its non-dimensional form. |
| $\lambda_e, \tilde{\lambda}_e$ | Measure of stretch due to external traction, Intrinsic kinematic variable, and its non-dimensional form. |
| $\lambda_p, \tilde{\lambda}_p$ | Measure of stretch due to fluidic pressure, Intrinsic kinematic variable, and its non-dimensional form. |
| $\alpha_s, \tilde{\alpha}_s$ | Total structure measure of curvature, Intrinsic kinematic variable, and its non-dimensional form. |
| $\alpha_e, \tilde{\alpha}_e$ | Measure of curvature due to external traction, Intrinsic kinematic variable, and its non-dimensional form. |
| $\alpha_p, \tilde{\alpha}_p$ | Measure of curvature due to fluidic pressure, Intrinsic kinematic variable, and its non-dimensional form. |
| $N_e, \tilde{N}_e$ | Cross sectional internal normal force resultant due to traction, and its non-dimensional form. |
| $V_e, \tilde{V}_e$ | Cross sectional internal shear force resultant due to traction, and its non-dimensional form. |
| $M_e, \tilde{M}_e$ | Cross sectional internal moment resultant due to traction, and its non-dimensional form. |
| $\boldsymbol{d^3}$ | The reciprocal vector to $\boldsymbol{d}_3$. |
| $\boldsymbol{b}, \boldsymbol{B}$ | External distributed force per unit mass. |
| $\boldsymbol{b^1}, \boldsymbol{B^1}$ | External distributed moment per unit mass. |
| $\boldsymbol{b_b}, \boldsymbol{B_b}$ | Body force distribution per unit mass, and its non-dimensional form. |
| $\boldsymbol{b_c}, \boldsymbol{B_c}$ | Contact force distribution per unit mass, and its non-dimensional form. |
| $\boldsymbol{b_b^1}, \boldsymbol{B_b^1}$ | First moment of body force distribution per unit mass, and its non-dimensional form. |
| $\boldsymbol{b_c^1}, \boldsymbol{B_c^1}$ | First moment of contact force distribution per unit mass, and its non-dimensional form. |
| $(\hat{\boldsymbol{x}}_f, \hat{\boldsymbol{y}}_f, \hat{\boldsymbol{z}}_f)$ | Fluidic domain curvilinear frame of reference. Defined such that the $\hat{\boldsymbol{x}}_f$ is the streamwise direction along bladder length $l_b$, and the cross sectional plane $\hat{\boldsymbol{y}}_f - \hat{\boldsymbol{z}}_f$ is perpendicular to $\hat{\boldsymbol{x}}_f$. |
| $(x_f, y_f, z_f)$ , $(X_f, Y_f, Z_f)$ | Fluid domain curvilinear coordinates, and their non-dimensional form. |
| $(u, v, w)$ , $(U, V, W)$ | Fluid domain velocity components, and their non-dimensional form. |
| $p, P$ | Fluid domain pressure, and its non-dimensional form. |
| $p', P'$ | Bladder effective fluid pressure for slope generation, and its non-dimensional form. |
| $\bar{p}, \bar{P}$ | Bladder effective fluid pressure for extension generation, and its non-dimensional form. |
| $q, Q$ | Volume flow rate in fluidic cross section, and its non-dimensional form. |
| $q_1^b, Q_1^b$ | Bladder permeability, and its non-dimensional form. |
| $q_1^c, Q_1^c$ | Connective tubes' permeability, and its non-dimensional form. |
| $a, A$ | Fluid domain cross section area, and its non-dimensional form. |
| $a_0, A_0$ | Cross section area of the bladder-tube array at gauge pressure $p = 0$, and its non-dimensional form. |
| $a_{p_1}, A_{p_1}$ | The change of the cross section area due to the fluid pressure, and its non-dimensional form. |
| $a_{N_1}, A_{N_1}$ | The change in cross section are due to extensional beam deformation, and its non-dimensional form. |
| $a_{M_1}, A_{M_1}$ | The change in cross section due to beam bending deformation, and its non-dimensional form. |

***Characteristic Scales***

|  |  |
|---|---|
| $u_1^*$ | Characteristic beam deflection. |
| $u_3^*$ | Characteristic beam extension. |
| $b_b^{1*}$ | Characteristic first moment of body force distribution per unit mass. |
| $b_c^{1*}$ | Characteristic first moment of contact force distribution per unit mass. |
| $b_b^*$ | Characteristic body force distribution per unit mass. |
| $b_c^*$ | Characteristic contact force distribution per unit mass. |
| $b_x^*$ | Characteristic shear force per unit length. |
| $b_z^*$ | Characteristic normal force per unit length. |
| $b^*$ | Characteristic external force per unit mass. |
| $b^{1*}$ | Characteristic external moment applied per unit mass. |
| $b_x^{1*}$ | Characteristic first moment applied by shear force $(x - axis)$ per unit length. |

| $b_z^{1*}$ | Characteristic first moment applied by normal force ($z - axis$) per unit length. |
|---|---|
| $\phi^*$ | Characteristic bladder density. |
| $t_s^*$ | Elastic-inertial time scale. |
| $\alpha_s^*$ | Characteristic curvature. |
| $M_e^*$ | Characteristic moment resultant. |
| $y_{11}^*$ | Characteristic squared radius of gyration. |
| $(u^*, v^*, w^*)$ | Characteristic fluid velocity. |
| $a_0^*$ | Characteristic cross-section at gage pressure. |
| $a_1^*$ | Characteristic change in bladder cross-section. |
| $t_f^*$ | Viscous-elastic time scale. |
| $u^*$ | Characteristic axial flow velocity. |
| $p^*$ | Characteristic fluid pressure. |

## A.1. Problem Formulation

We consider the dynamics of an elastic beam, initially at rest. The internal structure of the beam is a honeycomb-bladder matrix interconnected and filled with viscous fluid (as illustrated in Fig. A.1). Pressure within the fluid field both generates and is induced by the deformation of the solid liquid composite structure (denoted hereafter SLC).

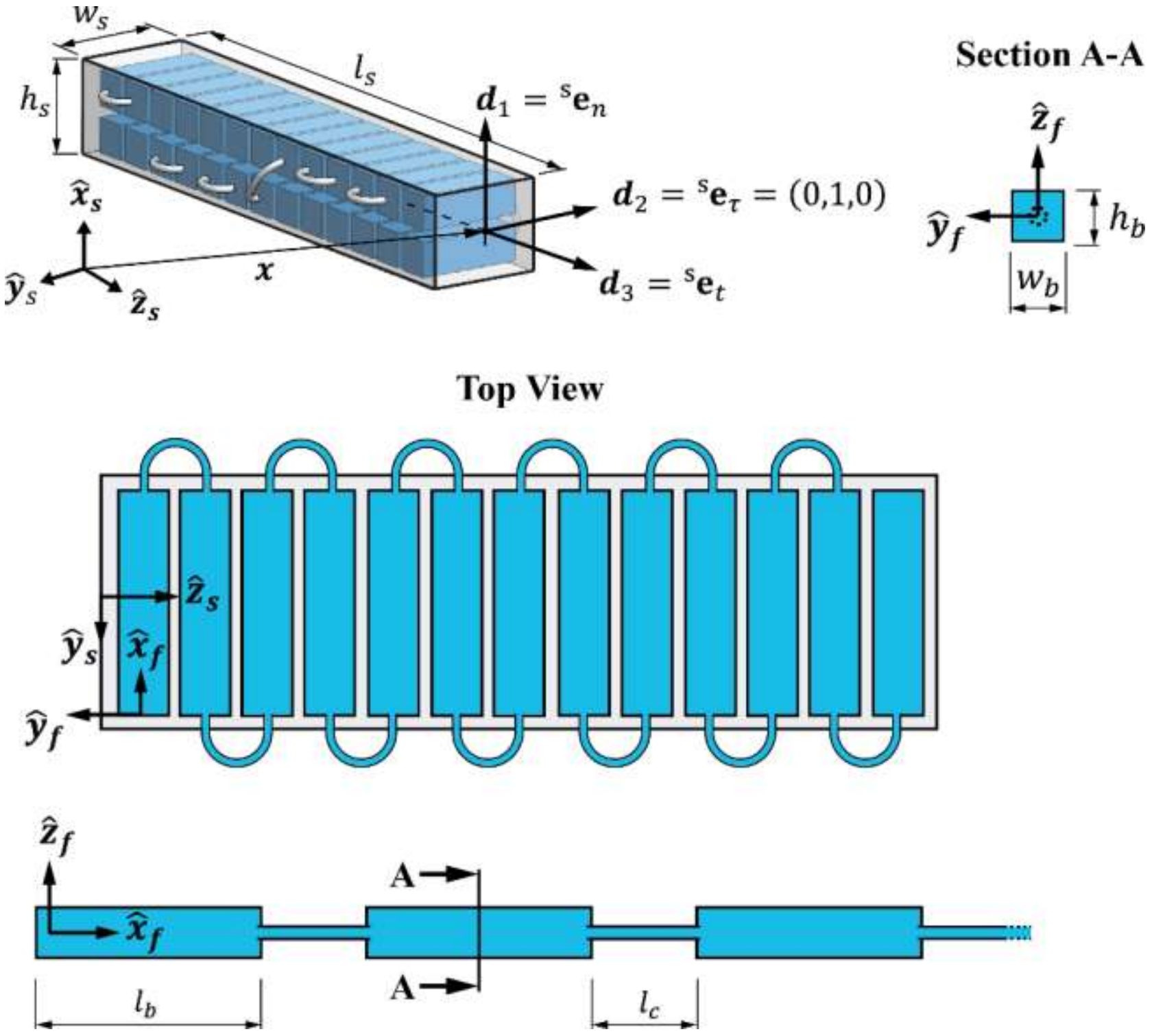

*FIG. A.1.* ***Illustration of an arbitrary Solid-Liquid-Composite (SLC) beam structure, with interconnected bladder-tube array in a honeycomb configuration.***

We define vector variables by bold letters, direction vectors by hat notation, non-dimensional variables by tilde or capital letters and characteristic values by asterisk superscripts. We define beam length $l_s$, beam height $h_s$, beam width $w_s$ and require a slender geometry with $h_s/l_s \ll 1$ and $w_s/l_s \ll 1$. Beam material modulus of elasticity, Poisson ratio and mass density are defined by $E, \nu$ and $\rho_s$, respectively. We define a lab frame of reference $(\boldsymbol{e}_1, \boldsymbol{e}_2, \boldsymbol{e}_3) = (\widehat{\boldsymbol{x}}_s, \widehat{\boldsymbol{y}}_s, \widehat{\boldsymbol{z}}_s)$ and a lab frame position vector $\boldsymbol{x} = (x_s(\theta,t), y_s(\theta,t),\ z_s(\theta,t))$ for the beam reference curve (i.e. the neutral axis). We define the deformation of a material fiber in the cross section area in a curvilinear frame of reference using the strain vectors, denoted spatial directors $(\boldsymbol{d}_1(\theta,t), \boldsymbol{d}_2(\theta,t), \boldsymbol{d}_3(\theta,t))$ and respective material directors $(\boldsymbol{D}_1(\theta), \boldsymbol{D}_2(\theta), \boldsymbol{D}_3(\theta))$. We define a Serret-Ferent triad associated with the current position along the reference curve in a curvilinear frame using the unit tangent ${}^s\boldsymbol{e}_t$ pointing the direction of motion, the unit normal ${}^s\boldsymbol{e}_n$ , unit binormal ${}^s\boldsymbol{e}_\tau$, and define a curvilinear length coordinate $\theta$ along the beam reference curve (see Figs. A.1).

We limit our analysis to a 2D deformation thus considering only directors $(\boldsymbol{d}_1(\theta,t), \boldsymbol{d}_3(\theta,t))$. We thus redefine the lab frame position vector $\boldsymbol{x} = (x_s(\theta,t), z_s(\theta,t))$ with the deflection axis in the $\boldsymbol{e}_1$ direction being $x_s(\theta,t) = x_0(\theta) + u_1(\theta,t)$ and extension axis in the $\boldsymbol{e}_3$ direction $z_s(\theta,t) = z_0(\theta) + u_3(\theta,t)$ where for a beam initially (at $t = 0$) straight at unstrained state, $x_0(\theta) = Const, z_0(\theta) = \theta$, and thus $x_s(\theta,t) = Const + u_1(\theta,t)$ and $z_s(\theta,t) = \theta + u_3(\theta,t)$ respectively.

The explicit representations of relevant directors are,

$$\boldsymbol{d}_1 = \left(\frac{\partial z_s}{\partial \theta}, -\frac{\partial x_s}{\partial \theta}\right), \tag{A.1}$$

$$\boldsymbol{d}_3 = \left(\frac{\partial x_s}{\partial \theta}, \frac{\partial z_s}{\partial \theta}\right). \tag{A.2}$$

An internal parallel honeycomb bladder matrix is interconnected and arranged perpendicular to the $\boldsymbol{d}_3$ direction along the beam length. The length of a single bladder segment is denoted $l_b$. The effect of the fluidic cavities on structure properties are represented by the solid mass fraction $f_m = ((w_b h_b l_b n)\rho_f + (l_s w_s h_s - w_b h_b l_b n)\rho_s)/(l_s w_s h_s \rho_s)$, mass per unit length $m = \rho_s(w_s h_s) f_m$, coefficients $f_e, f_i$ correct for cross section extensional and flexural stiffness reduction compared with a full elastic beam, beam cross section moment of inertia $I = (w_s h_s^3/12) f_i$ and the squared gyration radius $y^{11} = (\rho_s/m)I$. We limit our analysis to configurations with $w_b/l_s \ll 1$ and $(w_b \cdot n/2)/l_s \sim 1$, where $n$ is the total number of bladders, in order to approximate the above structural properties to constants representing an averaged property of the solid domain.

Constitutive laws are formulated using the intrinsic kinematic variables of $\lambda$ for the measure of stretch and $\alpha$ for curvature. The total stretch is defined by $\lambda_s = \lambda_e + \lambda_p$ and total curvature is defined by $\alpha_s = \alpha_e + \alpha_p$. Both $\lambda_e$ and $\alpha_e$ are due to external traction and $\lambda_p, \alpha_p$ are due to pressure in the fluidic cavities. We define the cross sectional internal forces and moment resultants due to traction for normal force $N_e$, shear force $V_e$ and moment $M_e$. A single pressurized bladder will create a change in beam slope $\psi$ and a change in beam length defined $\zeta$, and structure bladder density $\phi = (n/2)/l_s$.

We introduce a fluidic domain coordinate system $(x_f, y_f, z_f)$ defined such that the $\widehat{\boldsymbol{x}}_f$ is the streamwise direction along bladder length $l_b$ (see Fig. A.1). The plane $\widehat{\boldsymbol{y}}_f - \widehat{\boldsymbol{z}}_f$ is perpendicular to $\widehat{\boldsymbol{x}}_f$. Bladder height is $h_b$ and width $w_b$. We define a small parameter representing slenderness of the fluidic domain $\varepsilon_1 = 2r_c/l \ll 1$, where $r_c$ is the tube radius and $l = l_c\, n_c$ the total length of connective tubing, with $l_c$ the length of a single connective tube and $n_c$ as the total number of connective tubes in a given configuration. Tube and bladder characteristic cross section dimensional scale is $r_c \sim h_b$. The parameters of the fluidic domain are viscosity $\mu$, velocity $(u, v, w)$, gauge pressure $p$. Under small local strains assumption, bladder cross section area may be expanded to $a(x_f, p, N_e, M_e) = a_0(x_f) + a_{p_1}(p, x_f) + a_{N_1}(N_e, x_f) + a_{M_1}(M_e, x_f)$, where $a_0(x_f)$ is the cross section area of the fluidic domain i.e. bladder-tube array, at the gauge pressure $p = 0$, and $a_{p_1}(p, x_f)$ describes the change of the cross section area due to the fluid pressure, $a_{N_1}(N_e, x_f)$ the change in cross section are due to extensional beam deformation and $a_{M_1}(M_e, x_f)$ the change in cross section due to beam bending deformation. The governing equations for the incompressible, creeping, Newtonian flow are the stokes equation,

$$\boldsymbol{\nabla}\mathrm{p} = \mu\nabla^2\boldsymbol{u} \tag{A.3}$$

and conservation of mass

$$\boldsymbol{\nabla}\cdot\boldsymbol{u} = 0. \tag{A.4}$$

Over the solid domain we use an intrinsic Cosserat rod formulation following [55, 56], limited for the assumption of negligible cross sectional extension, cross sectional shear and tangential shear respectively limiting the cross section to maintain its initial shape and remain perpendicular the reference curve,

$$m\ddot{\boldsymbol{x}} = m\boldsymbol{b} + [N_{e},_{3} - \alpha_e V_e]^s\mathbf{e_t} + [V_{e},_{3} + \alpha_e N_e]^s\mathbf{e_n} + \left[\,\boldsymbol{d}^3\cdot m\left(\boldsymbol{b}^1 - y^{11}\,\ddot{\boldsymbol{d}}_1\,\right)\boldsymbol{d}_1\,\right],_3 \tag{A.5}$$

where $\boldsymbol{d}^3 = \boldsymbol{d}_3/d_{33}$ is the reciprocal vector to $\boldsymbol{d}_3$ and the subscript $,_3$ stands for the partial derivative with respect to $\theta$. The external distributed force $\boldsymbol{b}$ per unit mass, and external distributed moment per unit mass $\boldsymbol{b}^1$ are defined such,

$$\boldsymbol{b} = \boldsymbol{b}_b + \boldsymbol{b}_c\,, \tag{A.6}$$

$$\boldsymbol{b}^1 = \boldsymbol{b}_b^1 + \boldsymbol{b}_c^1, \tag{A.7}$$

where body force distribution per unit mass $\boldsymbol{b}_b = \left(b_{b_x}[Nm^2/Kg]\,, b_{b_z}[Nm^2/Kg]\right)\rho_s/m$, contact force distribution per unit mass $\boldsymbol{b}_c = \left(b_{c_x}[N/m], b_{c_z}[N/m]\right)/m$, first moment of body force distribution per unit mass $\boldsymbol{b}_b^1 = \frac{\rho_s}{m}\left(b_{b_x}^1[Nm^3/Kg]\,, b_{b_z}^1[Nm^3/Kg]\right)$, first moment of contact force distribution per unit mass $\boldsymbol{b}_c^1 = \frac{1}{m}\left(b_{c_x}^1[Nm/m]\,, b_{c_z}^1[Nm/m]\right)$. We define characteristic beam deflection $u_1^*[m]$, characteristic first moment of body force distribution per unit mass $b_b^{1*}[Nm/Kg]$, characteristic first moment of contact force distribution per unit mass $b_c^{1*}[Nm/Kg]$, characteristic body force distribution per unit mass $b_b^*[N/Kg]$, characteristic contact force distribution per unit mass $b_c^*[N/Kg]$, characteristic shear force per unit length $b_x^* \sim V_e^*/l_s[N/m]$, characteristic normal force per unit length $b_z^* \sim N_e^*/l_s[N/m]$, characteristic external force per unit mass $b^* \sim b_x^*/m[N/Kg]$, characteristic external moment applied per unit mass $b^{1*} \sim b_x^{1*}/m[Nm/Kg]$, characteristic first moment applied by shear force $(x - axis)$ per unit length $b_x^{1*} \sim b_x^*\,l_s[Nm/m]$, characteristic first moment applied by normal force $(z - axis)$ per unit length $b_z^{1*} \sim b_z^*\,l_s[Nm/m]$. Next, we denote the solid field characteristic beam extension $u_3^* \sim u_1^*$, characteristic bladder density $\phi^*[channel/m]$, characteristic elastic-inertial time scale $t_s^*[sec]$, characteristic curvature $\alpha_s^*[1/m]$, characteristic moment resultant $M_e^*[Nm]$, characteristic normal force resultant $N_e^*[N]$, characteristic shear force resultant $V_e^*[N]$ and characteristic squared radius of gyration $y_{11}^* \sim l_s^2[m^2]$. Over the fluid field, we define the characteristic velocity $(u^*, v^*, w^*)[m/sec]$, characteristic gauge pressure $p^*[Pa]$, characteristic fluidic domain cross-section at gage pressure $a_0^*[m^2]$, characteristic change in bladder cross-section $a_1^*[m^2]$ and viscous-elastic time scale $t_f^*[sec]$.

Next we define the normalized variables and coordinates. Normalized beam curvilinear coordinate $\Theta = \theta/l_s$, inertial-elastic time $T_s = t/t_s^*$, curvilinear deflection axis $X_s = x_s \backslash u_1^*$ and deflection variable $U_1 = u_1/u_1^* = u_1/l_s$, curvilinear extensional axis $Z_s = z_s/l_s$ and extension variable $U_3 = u_3/u_1^*$, beam curvature $\tilde{\alpha}_s$, beam stretch $\tilde{\lambda}_s$, moment resultant $\widetilde{M}_e = M_e/M_e^*$ , normal force resultant $\widetilde{N}_e = N_e/N_e^*$, shear force resultant $\tilde{V}_e = V_e/V_e^*$. First moment of body force distribution per unit mass $\boldsymbol{B}_b^1 = \Big(b_{b_x}^1/(\frac{b_x^{1^*}l_s^2}{m}), b_{b_z}^1/$ $(\frac{b_z^{1^*} l_s^2}{m})\Big)$, first moment of contact force distribution per unit mass $\boldsymbol{B}_c^1 = \left(b_{c_x}^1/b_x^{1^*}, b_{c_z}^1/b_z^{1^*}\right)$, body force distribution per unit mass $\boldsymbol{B}_b = \left(b_{b_x}/(\frac{b_x^* l_s^2}{m}), b_{b_z}/(\frac{b_z^* l_s^2}{m})\right)$, contact force distribution per unit mass $\boldsymbol{B}_c = \left(b_{c_x}/b_x^*, b_{c_z}/b_z^*\right)$. Fluidic domain spatial coordinates $\left(X_f, Y_f, Z_f\right) = \left(x_f/l,\ y_f/h_b,\ z_f/h_b\right)$, viscous-elastic time $T = t/t_f^*$, fluid velocity $(U, V, W) = (u/u^*,\ v/v^*,\ w/w^*)$, fluid field pressure $P = p/p^* = p/E$, bladder effective fluid pressure for slope generation $P' = p'/E$, bladder effective fluid pressure for extension generation $\bar{P} = \bar{p}/E$, volume flow rate in fluidic cross section $Q = q/(u^* a_0^*)$, bladder permeability $Q_1^b = q_1^b/\tilde{C}^b r_{eff}^4$ and connective tubes' permeability $Q_1^c = q_1^c/\tilde{C}^c r_{eff}^4$, where $r_{eff}$ and $\tilde{C}^i \sim 4\pi$ are respective effective scale and dimensionless constant related to the configuration of the flow-path i.e. shape of the cross-section, and $q_1^i$ $(i = c, b)$ is defined by the relation $q = -((1/\mu)\partial p/\partial x_f) q_1^i$. Fluidic cross section area is defined $a\left(x_f, p, N_e, M_e\right) = a_0(x_f) + a_{p_1}\left(p, x_f\right) + a_{N_1}\left(N_e, x_f\right) + a_{M_1}(M_e, x_f)$ and is normalized though $a_0^* = \pi r_c^2$ and $a_1^* = (\partial a_1/\partial p)p^*$ such that it reads $A(X_f, P, N_e, M_e) = A_0(X_f) + A_{p_1}\left(P, X_f\right)\sigma_p + A_{N_1}\left(N_e, X_f\right)\sigma_p + A_{M_1}\left(M_e, X_f\right)\sigma_p$ where $\sigma_p = a_1^*/a_0^*$. The slope introduced by a single bladder resulting from fluidic pressure and external traction are $\tilde{\alpha}_p = \alpha_p/\alpha_p^*$ and $\tilde{\alpha}_e = \alpha_e/\alpha_e^*$ respectively, extension introduced by a single bladder resulting from fluidic pressure and external traction are $\tilde{\lambda}_p = \lambda_p/\lambda_p^*$ and $\tilde{\lambda}_e = \lambda_e/\lambda_e^*$ respectively, non-dimensional squared gyration radius is $Y^{11} = y^{11}/y^{11*} = ((\rho_s/m) I f_i)/l_s^2$ and last bladder density along beam length $\Phi = \phi/\phi^* = \phi/((n/2)/l_s)$.

## A.2. Analysis

In this section we provide a formulation of the time-dependent model describing the two-way coupled physical mechanism of SLC systems. In §A.2.1 we present the formulation of the non-dimensional fluidic governing equations. In §A.2.2 we formulate the solid field Cosserat continuum constitutive laws. In §A.2.3 we introduce the Serret-Ferent triad associated with current position along the reference curve (natural axis). §A.2.4 we focuses on the formulation of the solid domain two scalar equations in the $X_s$ and $Z_s$ directions for the transverse $U_1$ and lateral $U_3$ deformation. In §A.2.5 we introduce the key step of formulating a two-way mapping between fluid and solid coordinate systems. In §A.2.6 presents the formulation of the bladder position identification function $R(x_f)$. In §A.2.7 we detail $\partial A_{N_1}\left(\widetilde{N}_e, \Theta\right)/\partial \widetilde{N}_e$, $\partial A_{M_1}\left(\widetilde{M}_e, \Theta\right)/\partial \widetilde{M}_e$, $\partial A_{p_1}(P, \Theta)/\partial P$ used to represent change in cross section due to section internal resultants $\widetilde{N}_e$, $\widetilde{M}_e$ and pressure $P$ respectively, as well as $Q_1\left(A\left(X_f, P\right)\right)$ the normalized volume flow rate. Last in §A.2.8, we introduce the methodology for separating our fluidic domain into required intervals with respect to our tubing configurations, and the application of boundary and initial conditions.

### A.2.1. Fluidic Field Governing Equations

Substituting the normalized variables into (A.3) and (A.4) yields in leading order,

$$\frac{\partial P}{\partial X_f} \sim \frac{\partial^2 U}{\partial {Y_f}^2} + \frac{\partial^2 U}{\partial {Z_f}^2}, \qquad \frac{\partial P}{\partial Y_f} \sim 0, \qquad \frac{\partial P}{\partial Z_f} \sim 0, \tag{A.8}$$

$$\frac{\partial U}{\partial X_f}+\frac{\partial V}{\partial Y_f}+\frac{\partial W}{\partial Z_f}\sim 0, \tag{A.9}$$

Where $2r_c/l \sim v^*/u^* = \varepsilon_1 \ll 1\mathrm{e}$ and $u^* = p^*\varepsilon_1^2 l/\mu$. Integrating (A.9) over the fluidic domain cross-section in the $Y_f - Z_f$ plane and applying Gauss theorem yields,

$$\frac{\partial Q}{\partial X_f}+\frac{h}{t_f^* v^*}\frac{\partial A}{\partial T}=0. \tag{A.10}$$

We define $Q_1\left(A(X_f,P)\right)$ as the normalized volume flow rate calculated by the solution of the Possion equation (A.8) for $\partial P/\partial X_f = -1$ with no-slip boundary condition set at the wall, $(U,V,W) = \boldsymbol{V}_{wall}$. From linearity, $Q$ can be obtained via $Q_1$ as

$$Q=-\frac{\partial P}{\partial X_f}Q_1\left(A(X_f,P)\right). \tag{A.11}$$

From order-of-magnitude analysis we obtain

$$q_1^{i*}=\tilde{C}^i\, r_{eff}^4\,,\qquad i=c,b \tag{A.12}$$

where $r_{eff}$ and $\tilde{C}^i \sim 4\pi$ are respectively the effective scale and dimensionless constant related to the configuration of the flow-path. Taking the derivative of $A(X_f,P,N_e,M_e) = A_0(X_f) + A_{p_1}(P,X_f)\sigma_p + A_{N_1}(N_e,X_f)\sigma_p + A_{M_1}(M_e,X_f)\sigma_p$ with regard to $T$ and substituting (A.11) into (A.10) we obtain,

$$\begin{aligned}
-&\left(\frac{\partial^2 P}{\partial {X_f}^2}\cdot Q_1\left(A(X_f,P)\right)+\frac{\partial P}{\partial X_f}\cdot\left(\frac{\partial Q_1}{\partial X_f}\right)\right)\\
&+\left(\frac{\partial A_{p_1}(P,\Theta)}{\partial P}\frac{\partial P}{\partial T}+\mathrm{R}(X_f)\frac{\partial A_{M_1}(\widetilde{M}_\mathrm{e},\Theta)}{\partial \widetilde{M}_e}\frac{\partial \widetilde{M}_e(\Theta,T)}{\partial T}\right.\\
&\left.+\left|\left(\mathrm{R}(X_f)\right)\right|\frac{\partial A_{N_1}(\widetilde{N}_\mathrm{e},\Theta)}{\partial \widetilde{N}_e}\frac{\partial \widetilde{N}_e(\Theta,T)}{\partial T}\right)=0.
\end{aligned} \tag{A.13}$$

The resulting non-linear diffusion equation represents the balance between the change in axial flux to the change of cross section area over time due to fluidic pressure, solid domain section moment resultant and normal force resultant. We define $R(X_f) = 1$ and $R(X_f) = -1$ discretely to indicate bladder position at the upper or lower row respectively. $\partial A_{M_1}/\partial \widetilde{M}_e$ represents the change in cross section due to section moment resultant and $\partial A_{N_1}/\partial \widetilde{N}_e$ for the change in cross section due to section normal force resultant. From order-of-magnitude analysis of (A.13) we obtain the viscous-elastic time scale $t_f^*$ as,

$$t_f^*=\frac{\sigma_p\mu}{p^*\varepsilon_1^2}=\frac{a_1^*\mu}{a_0^*p^*\varepsilon_1^2}. \tag{A.14}$$

For the case of small local strains of the fluidic cross section due to pressure, and for any bladder geometry, a proportional relation between $A_1$ and $P$ is upheld, and the viscous elastic time scale becomes $t_f^* = \mu(\partial a_1/\partial p)|_{p=p_0}/a_0^*\varepsilon_1^2$. This proportional relation is supported for a wide range of pressures as seen in the computation of $\partial a_{p_1}/\partial p$, see in §A.4.2.

Due to asymmetry of positive and negative change in cross section area resulting from moment and normal forces resultants. We define $\partial A_{M_1}/\partial \widetilde{M}_e$ as a piecewise function such that the proper coefficient is used per bladder position per moment sign,

$$\frac{\partial A_{M_1}}{\partial \widetilde{M}_e} = \left\{ \begin{array}{ll} R = 1 \quad \cap \quad \widetilde{M}_e \geq 0\,, & \dfrac{\partial A_{M_1}}{\partial \widetilde{M}_e} = \left(\dfrac{\partial A_{M_1}}{\partial \widetilde{M}_e}\right)_- \\ R = 1 \quad \cap \quad \widetilde{M}_e < 0\,, & \dfrac{\partial A_{M_1}}{\partial \widetilde{M}_e} = \left(\dfrac{\partial A_{M_1}}{\partial \widetilde{M}_e}\right)_+ \\ R = -1 \quad \cap \quad \widetilde{M}_e \geq 0\,, & \dfrac{\partial A_{M_1}}{\partial \widetilde{M}_e} = \left(\dfrac{\partial A_{M_1}}{\partial \widetilde{M}_e}\right)_+ \\ R = -1 \quad \cap \quad \widetilde{M}_e < 0\,, & \dfrac{\partial A_{M_1}}{\partial \widetilde{M}_e} = \left(\dfrac{\partial A_{M_1}}{\partial \widetilde{M}_e}\right)_- \end{array} \right\}, \tag{A.15}$$

and equivalently for $\partial A_1/\partial \widetilde{N}_e$ such that

$$\frac{\partial A_{N_1}}{\partial \widetilde{N}_e} = \left\{ \begin{array}{ll} \widetilde{N}_e \geq 0\,, & \dfrac{\partial A_{N_1}}{\partial \widetilde{N}_e} = \left(\dfrac{\partial A_{N_1}}{\partial \widetilde{N}_e}\right)_- \\ \widetilde{N}_e < 0\,, & \dfrac{\partial A_{N_1}}{\partial \widetilde{N}_e} = \left(\dfrac{\partial A_{N_1}}{\partial \widetilde{N}_e}\right)_+ \end{array} \right\} \tag{A.16}$$

where $\left(\partial A_{M_1}/\partial \widetilde{M}_e\right)_+$ and $\left(\partial A_{M_1}/\partial \widetilde{M}_e\right)_-$ represent the change in cross section due to section moment resultant that will induce a positive and negative pressure respectively, and $\left(\partial A_{N_1}/\partial \widetilde{N}_e\right)_+$ and $\left(\partial A_{N_1}/\partial \widetilde{N}_e\right)_-$ is the change in cross section due to section normal force resultant that will induce a positive and negative pressure respectively, see Fig. A.2. Of note is to mention, that due to the physical mechanism of positive and negative pressure generation,

$$\left(\frac{\partial A_{N_1}}{\partial \widetilde{N}_e}\right)_+ \geq \left(\frac{\partial A_{N_1}}{\partial \widetilde{N}_e}\right)_- > 0\,, \tag{A.17}$$

$$\left(\frac{\partial A_{M_1}}{\partial \widetilde{M}_e}\right)_+ \geq \left(\frac{\partial A_{M_1}}{\partial \widetilde{M}_e}\right)_- > 0\,. \tag{A.18}$$

Equation (A.13) requires two boundary conditions and one initial condition to be set per fluidic domain interval, see §A.2.8.

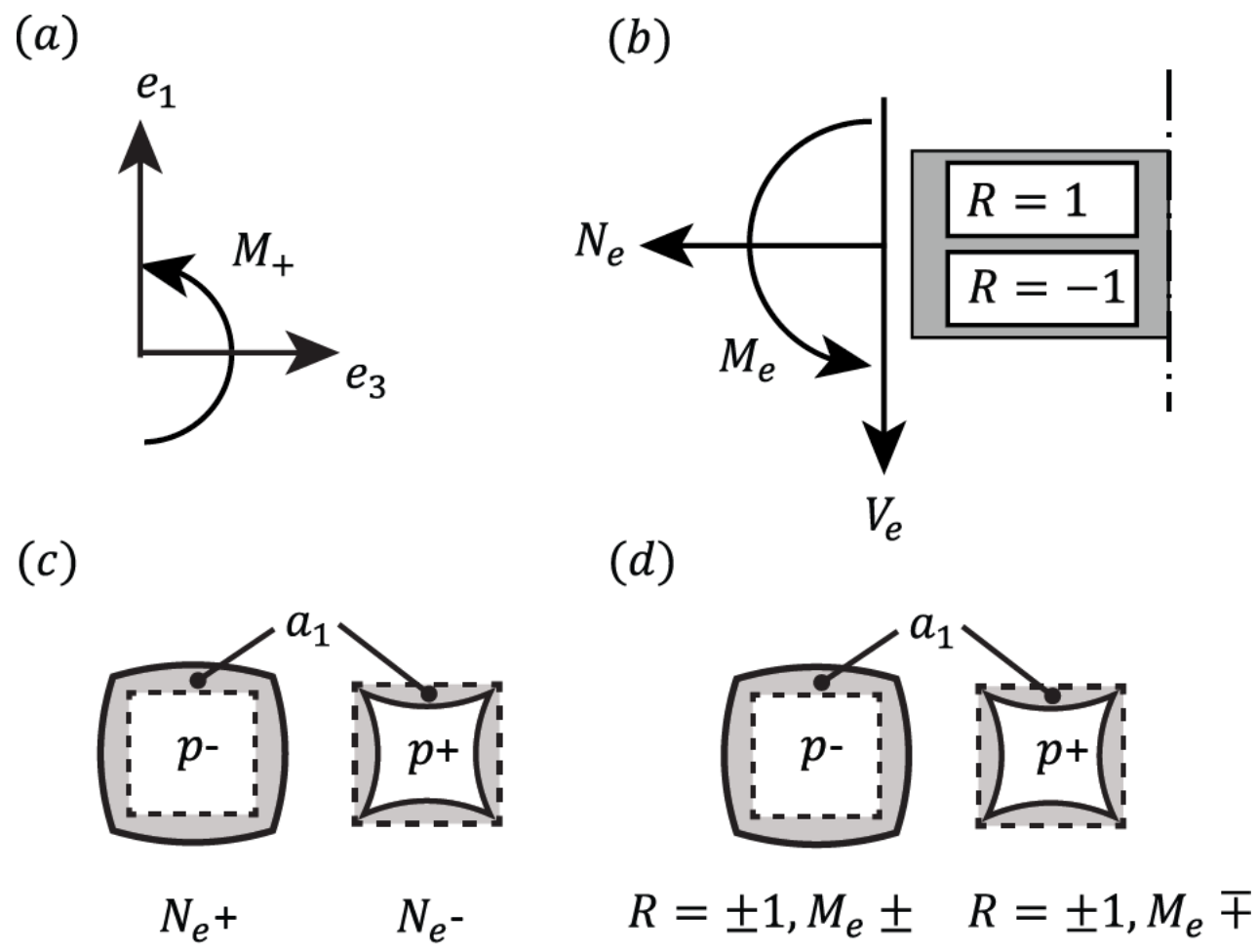

*FIG. A.2.* ***Illustration of Sign convention using right-handed system.*** *(a) Lab-frame and positive moment defined. (b) Illustration of section internal/resultant forces and moment* $\mathrm{V_e}, \mathrm{N_e}, \mathrm{M_e}$. *Bladder change in cross section* $\mathrm{a_1}$ *presented in response to section resultants: (c)* $\mathrm{N_e}$ *and (d)* $\mathrm{M_e}$. *Bladder section initial area* $\mathrm{a_0}$, *is denoted in dashed lines, gray areas indicate induced change in cross section* $\mathrm{a_1}$ *due to respective resultant.*

### A.2.2. Constitutive Laws for a Two-Way Coupled SLC

We now turn formulate the solid field Cosserat continuum constitutive laws. Our intrinsic kinematic variables $\lambda_s[1] = \lambda_e + \lambda_p$ stands for the total stretch measure and $\alpha_s[1/m] = \alpha_e + \alpha_p$ for the total curvature measure of the reference curve. The subscript $e$ and $p$ represent the source of the measure being external forces i.e. traction applied to the surface, and pressure respectively. We define in dimensional form,

***Inrinsic Kinematic Variables***

$$\lambda_s = \frac{{}^s d_{33}^{1/2}}{{}^s D_{33}^{1/2}}, \tag{A.19}$$

$$\alpha_s = \frac{{}^s \boldsymbol{d}_{1,3} \cdot {}^s \boldsymbol{d}_3}{{}^s d_{33}^{1/2}\, {}^s D_{33}^{1/2}}, \tag{A.20}$$

where $d_{33} = \boldsymbol{d}_3 \cdot \boldsymbol{d}_3$ defines the metric's of $\boldsymbol{d}_3$ vector at present configuration i.e. spatial frame, and $D_{33} = \boldsymbol{D}_3 \cdot \boldsymbol{D}_3$ defining the metric's of $\boldsymbol{d}_3$ vector at $t = 0$ configuration i.e. material frame. For a beam (straight rod) at relaxed state oriented along the lab frame results in $(\boldsymbol{D}_1, \boldsymbol{D}_2, \boldsymbol{D}_3) = (\boldsymbol{e}_1, \boldsymbol{e}_2, \boldsymbol{e}_3)$, thus we can formulate

$$D_{33} = \boldsymbol{D}_3 \cdot \boldsymbol{D}_3 = \boldsymbol{e}_3 \cdot \boldsymbol{e}_3 = (0,0,1) \cdot (0,0,1) = 1. \tag{A.21}$$

The pressure induced measures for stretch and curvature are defined

$$\lambda_p = \frac{\bar{p}}{E} \frac{\partial \lambda_p}{\partial(\bar{p}/E)}, \tag{A.22}$$

$$\alpha_p = -\frac{p'}{E} \cdot \frac{\partial \alpha_p}{\partial(p'/E)} \tag{A.23}$$

where the effective pressure for slope and extension generation are respectively $p' = (p_d - p_u)$ and $\bar{p} = (p_d + p_u)$. where $p_u$ and $p_d$ are the fluidic pressures at the upper and lower bladders, $\partial\lambda_p/\partial(\bar{p}/E)$ is the measure of stretch per cross section per normalized pressure sum and $\partial\alpha_p/\partial(p/E)$ is the measure of curvature per cross section per normalized pressure difference. Formulating the respective non-dimensional form yields

$$\frac{\partial\tilde{\alpha}_p}{\partial(p'/E)} \approx \Phi\frac{\partial\psi}{\partial(p'/E)}, \tag{A.24}$$

$$\frac{\partial\tilde{\lambda}_p}{\partial(\bar{p}/E)} = \Phi\frac{\partial\tilde{\zeta}}{\partial(\bar{p}/E)} \tag{A.25}$$

where $\partial\psi/\partial(p'/E)$ represents the change in beam slope per cross section per normalized pressure difference and $\partial\tilde{\zeta}/\partial(\bar{p}/E)$ the non-dimensional change in length per cross section per normalized pressure sum. The later $\partial\tilde{\zeta}/\partial(\bar{p}/E)$ may now be formulated in dimensional in form

$$\frac{\partial\tilde{\zeta}}{\partial(\bar{p}/E)} = \phi^*\frac{\partial\zeta}{\partial(\bar{p}/E)} \tag{A.26}$$

where $\partial\zeta/\partial(\bar{p}/E)$ is the dimensional change in length per cross section per normalized pressure sum.

Substituting our normalized variables into equations (A.1), (A.2) and (A.19) - (A.23) respectively, we obtain our directors and kinematic variables in non-dimensional form,

$$\tilde{\boldsymbol{d}}_1 = \left(\left(\frac{l_s}{u_1^*}\right) + \frac{\partial U_3}{\partial\Theta}, -\frac{\partial U_1}{\partial\Theta}\right), \tag{A.27}$$

$$\tilde{\boldsymbol{d}}_3 = \left(\frac{\partial U_1}{\partial\Theta}, \left(\frac{l_s}{u_1^*}\right) + \frac{\partial U_3}{\partial\Theta}\right) \tag{A.28}$$

and,

$$\tilde{\lambda}_e = \frac{1}{\lambda_e^*}\underbrace{\frac{{}^s\tilde{d}_{33}^{1/2}}{{}^s\tilde{D}_{33}^{1/2}}}_{\tilde{\lambda}_s} - \frac{\lambda_p^*}{\lambda_e^*}\underbrace{\bar{P}(X_f)\frac{\partial\tilde{\lambda}_p}{\partial(\bar{p}/E)}}_{\tilde{\lambda}_p}, \tag{A.29}$$

$$\tilde{\alpha}_e = \frac{1}{l_s\alpha_e^*}\underbrace{\frac{\tilde{d}_{1x,3}\,\tilde{d}_{3x} + \tilde{d}_{1z,3}\,\tilde{d}_{3z}}{{}^s\tilde{d}_{33}^{1/2}\,{}^s\tilde{D}_{33}^{1/2}}}_{\tilde{\alpha}_s} + \frac{\alpha_p^*}{\alpha_e^*}\underbrace{P'(X_f)\frac{\partial\tilde{\alpha}_p}{\partial(p'/E)}}_{\tilde{\alpha}_p}. \tag{A.30}$$

From Order-of-magnitude analysis of we determine the characteristic scale for the directors $d_1^* \sim u_1^*/l_s$ and $d_3^* = u_1^*/l_s$, as well as for the measure of curvature $\alpha_e^* \sim \alpha_p^* \sim \frac{1}{l_s}$ and stretch $\lambda_e^* \sim \lambda_p^* \sim 1$.

***Constitutive Equations, Force and Moment Resultants***

The constitutive equations for normal force resultant $N_e$, shear force resultant $V_e$ and bending moment resultant $M_e$ are now formulated in dimensional form,

$$N_e = Eh_s w_s f_e(\lambda_e - 1), \tag{A.31}$$

$$M_e = E\frac{w_s h_s^3}{12} f_i \cdot \alpha_e, \tag{A.32}$$

$$V_e = -{}^s d_{33}^{-1/2} M_{e,3}. \tag{A.33}$$

Substituting the normalized variables (A.31)-(A.33) become

$$\widetilde{N}_e = \left(\tilde{\lambda}_e - \frac{1}{\lambda_e^*}\right), \tag{A.34}$$

$$\widetilde{M}_e = \tilde{\alpha}_e , \tag{A.35}$$

$$\tilde{V}_e = -\frac{1}{{}^s\tilde{d}_{33}^{1/2}}\frac{\partial \widetilde{M}_e}{\partial \Theta}. \tag{A.36}$$

With order-of-magnitude analysis respectively yielding,

$$N_e^* \sim E h_s w_s f_e \lambda_e^* , \tag{A.37}$$

$$M_e^* \sim E\frac{w_s h_s^3}{12} f_i \frac{1}{l_s}, \tag{A.38}$$

$$V_e^* \sim E\frac{w_s h_s^3}{12} f_i \frac{1}{l_s^2}. \tag{A.39}$$

## A.2.3. Serret-Ferent Curvilinear Coordinates

The Serret-Ferent triad associated with the current position along the reference curve is characterized by the unit tangent ${}^s\boldsymbol{e}_t$ along beam length, the unit normal ${}^s\boldsymbol{e}_n$ and unit binormal ${}^s\boldsymbol{e}_\tau$, see illustrated in Fig. A.1. In dimensional form,

$${}^s\mathbf{e_t} = \frac{{}^s\boldsymbol{d}_3(\theta, t)}{{}^s d_{33}^{1/2}}, \tag{A.40}$$

$${}^s\mathbf{e_n} = -\frac{({}^s\boldsymbol{d}_{1,3} \cdot {}^s\boldsymbol{d}_3){}^s\boldsymbol{d}_1}{\alpha_s \lambda_s {}^s D_{33}}. \tag{A.41}$$

Substituting the normalized variables yields,

$${}^s\boldsymbol{E_t} = \frac{\left(\tilde{d}_{3x}, \tilde{d}_{3z}\right)}{{}^s\tilde{d}_{33}^{1/2}}, \tag{A.42}$$

$${}^s\boldsymbol{E_n} = -\frac{\left(\tilde{d}_{1x}, \tilde{d}_{1z}\right)}{{}^s\tilde{d}_{33}^{\frac{1}{2}}}. \tag{A.43}$$

## A.2.4. Solid Field Governing Equations

For the two way coupled solid field governing equations, the intrinsic Cosserat rod formulation is used with both the deflection component $U_1$ in the $\boldsymbol{e}_1$ lab frame direction and tangential deformation component $U_3$ in the $\boldsymbol{e}_3$ direction included. Substituting (A.6), (A.7), (A.19) - (A.43) in conjunction with normalized variables and applying order of magnitude analysis onto (A.5) we obtain two scalar equation; one in the $X_s$ direction

$$\begin{aligned} X_s\colon \Pi_1\left(\frac{\partial^2 U_1}{\partial T^2}\right) &= B_x + \left[\Pi_2 \frac{\partial \widetilde{N}_e}{\partial \Theta} - \Pi_3 \tilde{\alpha}_e \tilde{V}_e\right]{}^s E_{tx} + \left[\Pi_4 \frac{\partial \tilde{V}_e}{\partial \Theta} + \Pi_5 \tilde{\alpha}_e \widetilde{N}_e\right]{}^s E_{nx} \\ &+ \frac{\partial}{\partial \Theta}\left[\frac{1}{\tilde{d}_{33}}\left(\left(\Pi_6 B_x^1 - \Pi_7 Y^{11}\frac{\partial^2 \tilde{d}_{1x}}{\partial T^2}\right)\tilde{d}_{3x}\right.\right. \\ &\left.\left. + \left(\Pi_8 B_z^1 - \Pi_9 Y^{11}\frac{\partial^2 \tilde{d}_{1z}}{\partial T^2}\right)\tilde{d}_{3z}\right)\tilde{d}_{1x}\right] \end{aligned} \tag{A.44}$$

and in the $Z_s$ direction

$$
\begin{aligned}
Z_s: \mathit{\Pi}_1\left(\frac{\partial^2 U_3}{\partial T^2}\right) = B_z &+ \left[\mathit{\Pi}_2 \frac{\partial \widetilde{N}_e}{\partial \Theta} - \mathit{\Pi}_3 \tilde{\alpha}_e \widetilde{V}_e\right] {}^sE_{tz} + \left[\mathit{\Pi}_4 \frac{\partial \widetilde{V}_e}{\partial \Theta} + \mathit{\Pi}_5 \tilde{\alpha}_e \widetilde{N}_e\right] {}^sE_{nz} \\
&+ \frac{\partial}{\partial \Theta}\left[\frac{1}{\tilde{d}_{33}}\left(\left(\mathit{\Pi}_6 B_x^1 - \mathit{\Pi}_7 Y^{11} \frac{\partial^2 \tilde{d}_{1x}}{\partial T^2}\right) \tilde{d}_{3x}\right.\right. \\
&\left.\left. + \left(\mathit{\Pi}_8 B_z^1 - \mathit{\Pi}_9 Y^{11} \frac{\partial^2 \tilde{d}_{1z}}{\partial T^2}\right) \tilde{d}_{3z}\right) \tilde{d}_{1z}\right]
\end{aligned}
\tag{A.45}
$$

where both are scaled by non-dimensional numbers $\mathit{\Pi}_1 = \left(t_s^*/t_f^*\right)^2$, $\mathit{\Pi}_2 = N_e^*/(l_s m b^*)$ , $\mathit{\Pi}_3 = (\alpha_e^* V_e^*)/(mb^*)$ , $\mathit{\Pi}_4 = V_e^*/(l_s m b^*)$ , $\mathit{\Pi}_5 = (\alpha_e^* N_e^*)/(mb^*)$ , $\mathit{\Pi}_6 = \mathit{\Pi}_8 = (d_1^* b^{1*} d_3^*)/(l_s d_{33}^* b^*)$ , $\mathit{\Pi}_7 = \mathit{\Pi}_9 = \left((d_1^* y^{11*} d_3^*)/(l_s^2\, d_{33}^*)\right)\left(t_s^*/t_f^*\right)^2$, and the force and first moment distribution per unit mass respectively defined,

$$
\boldsymbol{B} = \left(\frac{\rho_s l_s^2}{m} B_{b_x} + B_{c_x} \ , \ \frac{\rho_s l_s^2}{m}\left(\frac{b_z^*}{b_x^*}\right) B_{b_z} + \left(\frac{b_z^*}{b_x^*}\right) B_{c_z}\right),
\tag{A.46}
$$

$$
\boldsymbol{B}^1 = \left(\frac{\rho_s l_s^2}{m} B_{b\,x}^1 + B_{c\,x}^1 \ , \ \frac{\rho_s l_s^2}{m}\left(\frac{b_z^{1^*}}{b_x^{1^*}}\right) B_{b\,z}^1 + \left(\frac{b_z^{1^*}}{b_x^{1^*}}\right) B_{c\,z}^1\right).
\tag{A.47}
$$

Order of magnitude analysis of $\boldsymbol{B}, \boldsymbol{B}^1$ yields $b_x^* \sim V_e^*/l_s$, $b^* \sim b_x^*/m$, $b_x^{1*} \sim b_x^* l_s$, $b^{1*} \sim b_x^{1^*}/m$, $b_z^* \sim N_e^*/l_s$, $b_z^{1*} \sim b_z^* l_s$, $y^{11*} \sim l_s^2$. Determining the relevant dynamic regime of the SLC structure we define the time scale ratio

$$
\frac{t_s^*}{t_f^*} = \frac{a_0^* \varepsilon_1^2 \sqrt{m u_1^* l_s^3/EIf_i}}{\mu(\partial a_1/\partial p)|_{p=p_0}}
\tag{A.48}
$$

where $t_s^* \sim \sqrt{m u_1^* l_s^3/EIf_i}$ is the elastic-inertial time-scale and $t_f^* = \mu(\partial a_1/\partial p)|_{p=p_0}/a_0^* \varepsilon_1^2$ the viscous-elastic time scale.

Equations (A.44) and (A.45) require six boundary conditions and four initial conditions, with $\mathrm{K}_i$ as the boundary or initial condition function. We define possible boundary conditions as a set of geometric and dynamic conditions. Geometric conditions are applied in similar methodology as is the case for a classic Euler-Bernoulli beam where boundary conditions are applied over the overall structure quantity of the reference curve. Thus over $U_1$ for deflection

$$
U_1(\gamma, T) = \mathrm{K}_1(T)
\tag{A.49}
$$

or slope

$$
\left.\left(\frac{\partial U_1}{\partial \Theta}\right)\right|_{(\gamma,T)} = \mathrm{K}_2(T)\,,
\tag{A.50}
$$

and for $U_3$ extension,

$$
U_3(\gamma, T) = \mathrm{K}_3(T)
\tag{A.51}
$$

where $\gamma = 0$ or $\gamma = 1$ are used to set conditions at respective boundaries. Dynamic conditions on the other hand, relate to additional displacement due to external moments and normal, shear forces at the boundary. Using (A.30), (A.35) and (A.36), Dynamic conditions are thus applied for moment

$$
\left.\widetilde{M}_e\right|_{(\gamma,T)} = \mathrm{K}_4(T)\,,
\tag{A.52}
$$

shear force

$$\tilde{V}_e\Big|_{(\gamma,T)} = \mathrm{K}_5(T)\,, \tag{A.53}$$

and respectively for normal forces over $U_3$ using (A.34) and (A.29)

$$\tilde{N}_e\Big|_{(\gamma,T)} = \mathrm{K}_6(T)\,. \tag{A.54}$$

Last, initial conditions are directly applied over $U_1$ and $U_3$

$$U_1(\Theta,0) = \mathrm{K}_7(\Theta) \tag{A.55}$$

and

$$U_3(\Theta,0) = \mathrm{K}_8(\Theta)\,. \tag{A.56}$$

As well as over the initial time derivative,

$$\left(\frac{\partial U_1}{\partial T}\right)\Bigg|_{(\Theta,0)} = \mathrm{K}_9(\Theta) \tag{A.57}$$

and

$$\left(\frac{\partial U_3}{\partial T}\right)\Bigg|_{(\Theta,0)} = \mathrm{K}_{10}(\Theta)\,. \tag{A.58}$$

### A.2.5. Coordinate Mapping Between Solid and Fluid Domains

A key step in the formulating of the model of a given system is a two-way mapping of the fluid field onto the solid. Coordinate mapping has a dual purpose: first, to correlating local force and moment resultants to respective regions in the fluid field and vice versa, second, to formulate the piecewise-continuous fluid field along which pressure diffuses respective to bladder tubing configuration. Fig. A.3 shows the schematic setup of three systems utilized in this article, upper serpentine (panel a), parallel serpentine configuration (panel b) and crossover configuration (panel c). Principles of coordinate mapping presented herein are illustrated on the selected systems' configuration. In this section we present an algorithm generalized for any arbitrary configuration. We define our bladder index $j$ by order of geometric position along $\Theta$ such that $j \in [1, n/2]$ from left to right per row, table row index $k \in [1, n]$. We denote index $\mathscr{s}$ holding in successive order the $(R \cdot j)$ index of connected bladders by order of connection not position in the connectivity array $\Omega_c$ i.e. for the Upper serpentine configuration $\Omega_\mathrm{c} = \{1,2,3,4,5,6,7,8,9,10\}$ , for parallel *serpentine* configuration $\Omega_\mathrm{c} = \{1,2,3,4,5,6,7,8,9,10\}$ *&* $\{-1,-2,-3,-4,-5,-6,-7,-8,-9,-10\}$ *and* last for crossover configuration $\Omega_\mathrm{c} = \{1,2,3,-4,-5,-6,-7,-8,-9,-10\}$ *&* $\{-1,-2,-3,4,5,6,7,9,9,10\}$. We denote index $\mathscr{q}$ holding in successive order the $(R \cdot j)$ index of disconnected bladders by order of left-to-right and top-to-bottom for the discontinuous array $\Omega_d$ , i.e. for *upper serpentine configuration* $\Omega_\mathrm{d} = \{-1,-2,-3,-4,-5,-6,-7,-8,-9,-10\}$, for parallel serpentine configuration $\Omega_d = \{\emptyset\}$ and for *crossover configuration* $\Omega_\mathrm{d} = \{\emptyset\}$.

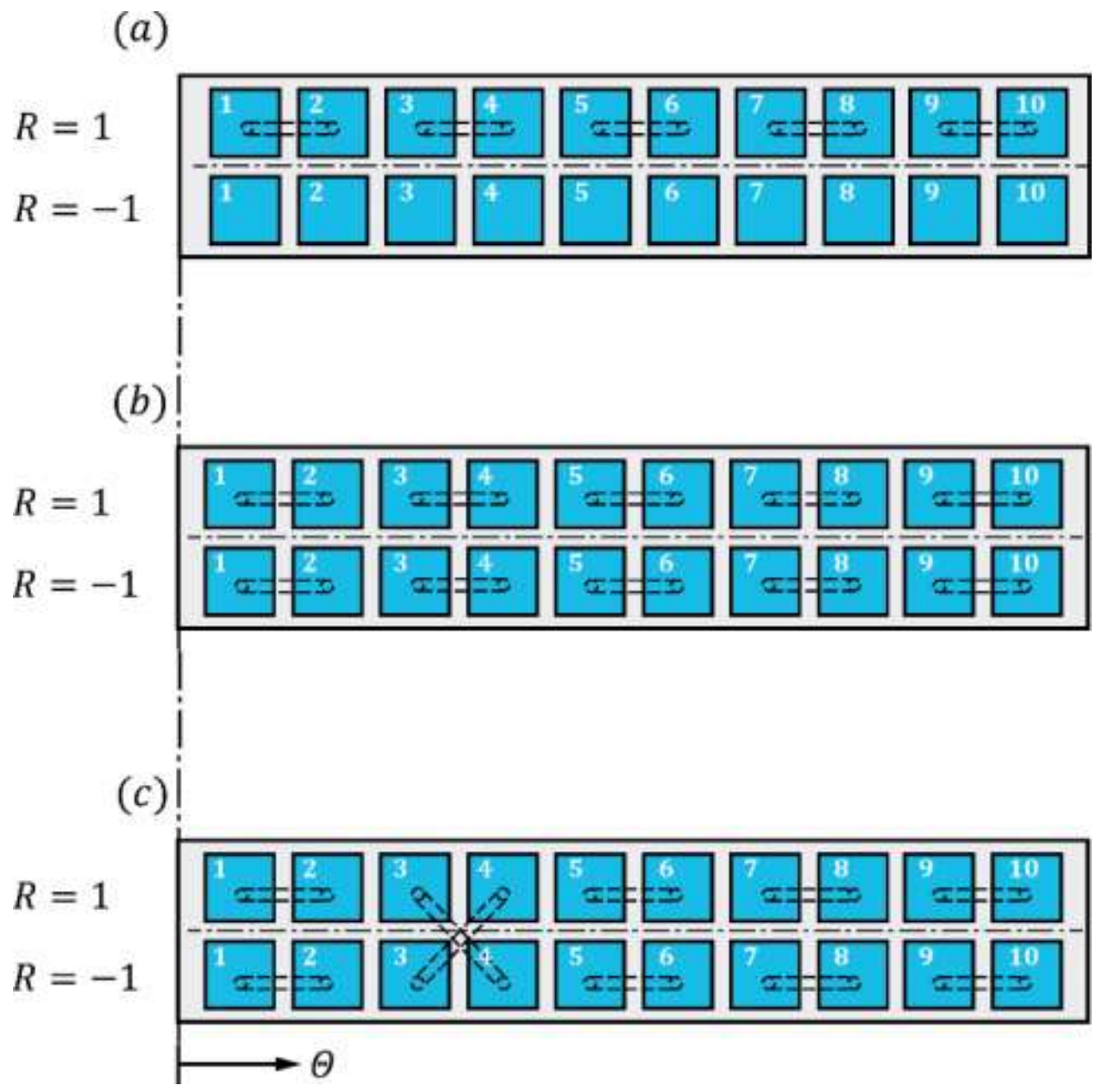

*FIG. A.3.* ***Illustration of a schematic beam section with top and bottom bladder connected in various configurations.*** *(a) Upper serpentine* $\Omega_{\mathrm{c}} = \{1,2,3,4,5,6,7,8,9,10\}$ , $\Omega_{\mathrm{d}} = \{-1,-2,-3,-4,-5,-6,-7,-8,-9,-10\}$ . *(b) Parallel serpentine* $\Omega_{\mathrm{c}} = \{1,2,3,4,5,6,7,8,9,10\}$ *&* $\{-1,-2,-3,-4,-5,-6,-7,-8,-9,-10\}$ , $\Omega_{\mathrm{d}} = \{\emptyset\}$ . *(c) Crossover* $\Omega_{\mathrm{c}} = \{1,2,3,-4,-5,-6,-7,-9,-9,-10\}$ *&* $\Omega_{\mathrm{c}} = \{-1,-2,-3,4,5,6,7,8,9,10\}$, $\Omega_{\mathrm{d}} = \{\emptyset\}$. *Bladders' index* j *Marked in white per bladder.*

***$\Theta - X_f$ Coordinate Mapping, $X_f(\Theta)$***

We start by mapping the fluid pressure onto the solid field. In equations (A.44) and (A.45) pressure governs a change in slope and extension source terms $\tilde{\alpha}_p$ and $\tilde{\lambda}_p$ by $P'(X_f)$ and $\bar{P}(X_f)$ respectivly. We thus need to map our pressure onto the solid field to produce their respective value. Table column $(\Theta \cdot R)$ represents the non-dimensional coordinate along beam length, where $\Theta \in [0,1]$, $R = 1$ is the index upper row bladders and $R = -1$ for lower respectively. The mapping of beam length position to lower row bladders is mapped onto the range $(\Theta \cdot R) < 0$ and for upper row bladders over $(\Theta \cdot R) > 0$. Size of a single bladder along the $\Theta$ coordinate for the purpose of coordinate mapping is the total non-dimensional length $\Theta = 1$ divided by the number of bladders in one row $|(\Theta \cdot R)| = (1/(\mathrm{n}/2))$. We define a small coordinate length parameter $\epsilon = o(1/(n/2))$. The $(\Theta \cdot R)$ column is thus

$$(\Theta \cdot R)_{2k-1} = \begin{Bmatrix} k \leq \left(\frac{n}{2}+1\right), & \left(-\left(\frac{n}{2}\right)+(k-1)\right)\cdot\left(\frac{n}{2}\right)^{-1} \\ k > \left(\frac{n}{2}+1\right), & \left(-\left(\frac{n}{2}\right)+(k-1)\right)\cdot\left(\frac{n}{2}\right)^{-1}+\epsilon \end{Bmatrix}$$
$$(\Theta \cdot R)_{2k} = \begin{Bmatrix} k \leq \left(\frac{n}{2}\right), & \left(-\left(\frac{n}{2}\right)+k\right)\cdot\left(\frac{n}{2}\right)^{-1}-\epsilon \\ k > \left(\frac{n}{2}\right), & \left(-\left(\frac{n}{2}\right)+k\right)\cdot\left(\frac{n}{2}\right)^{-1} \end{Bmatrix}, \tag{A.59}$$

these coordinates represent the start and end points of each bladder along the $(\Theta \cdot R)$ coordinate. Next we address the $X_f$ column. Setting our connective array parameters $n$, $n_c$, $l_c$, $l_b$, we calculate the total length of bladder-tube array for mapping purposes

$$\ell_{tot} = l_b n + l_c n_c \,. \tag{A.60}$$

Solid field coordinate $\Theta$ maps to the center position along the bladder length $l_b$. We define the non-dimensional bladder and tube lengths $\mathcal{L}_b = l_b/\ell_{tot}$ and $\mathcal{L}_c = l_c/\ell_{tot}$ respectively, and construct the $X_f$ mapped position such that connected bladders are addressed first; The first connected bladder is set $X_f = \mathcal{L}_b/2$ spanning a length of $(\Theta \cdot R)$ of respective index $(R \cdot j)$. Consecutive-connected bladders are then added respectively in table rows; each $X_f^{s}$ bladder coordinate spans a length of $(\Theta \cdot R)$ respective to its bladder index$(R \cdot j)$ i.e. the coordinate value $X_f$ appears in two rows representing the length of the beam $\Theta$ that houses the bladder of index $(R \cdot j)$. The $X_f$ position of consecutive bladders is $X_f^{s} = X_f^{s-1} + \mathcal{L}_b + \mathcal{L}_c$. In the case of separate interconnected bladder arrays, the $X_f$ position for the first bladder in the new array is $X_f^{s} = X_f^{s-1} + \mathcal{L}_b$ where $X_f^{s-1}$ is the coordinate value from the last connected bladder of the previous array. We then proceed as detailed above for consecutive connected bladders until we finish all interconnected arrays. Next, we turn to address the disconnected bladders by order of $\Omega_d$; each bladder starts at its initial position $(\Theta \cdot R)$ respective to its bladder index $(R \cdot j)$ and spans the length $(\Theta \cdot R)$ of that index. In the $X_f$ dimension, $X_f^{q} = X_f^{q-1} + \mathcal{L}_b$ where for the first disconnected bladder's $X_f^{q-1}$ is the coordinate value from the last connected bladder. The mapping detailed above generates a $C^0$ continuous mapping function. Fig. A.4 illustrates the plot of coordinate mappings' for three key examples, upper serpentine (panel a), parallel serpentine (panel b) and crossover (panel c). Respective tabulated data used in plots is presented in Table A.1a, A.1b and A.1c in §A.5.

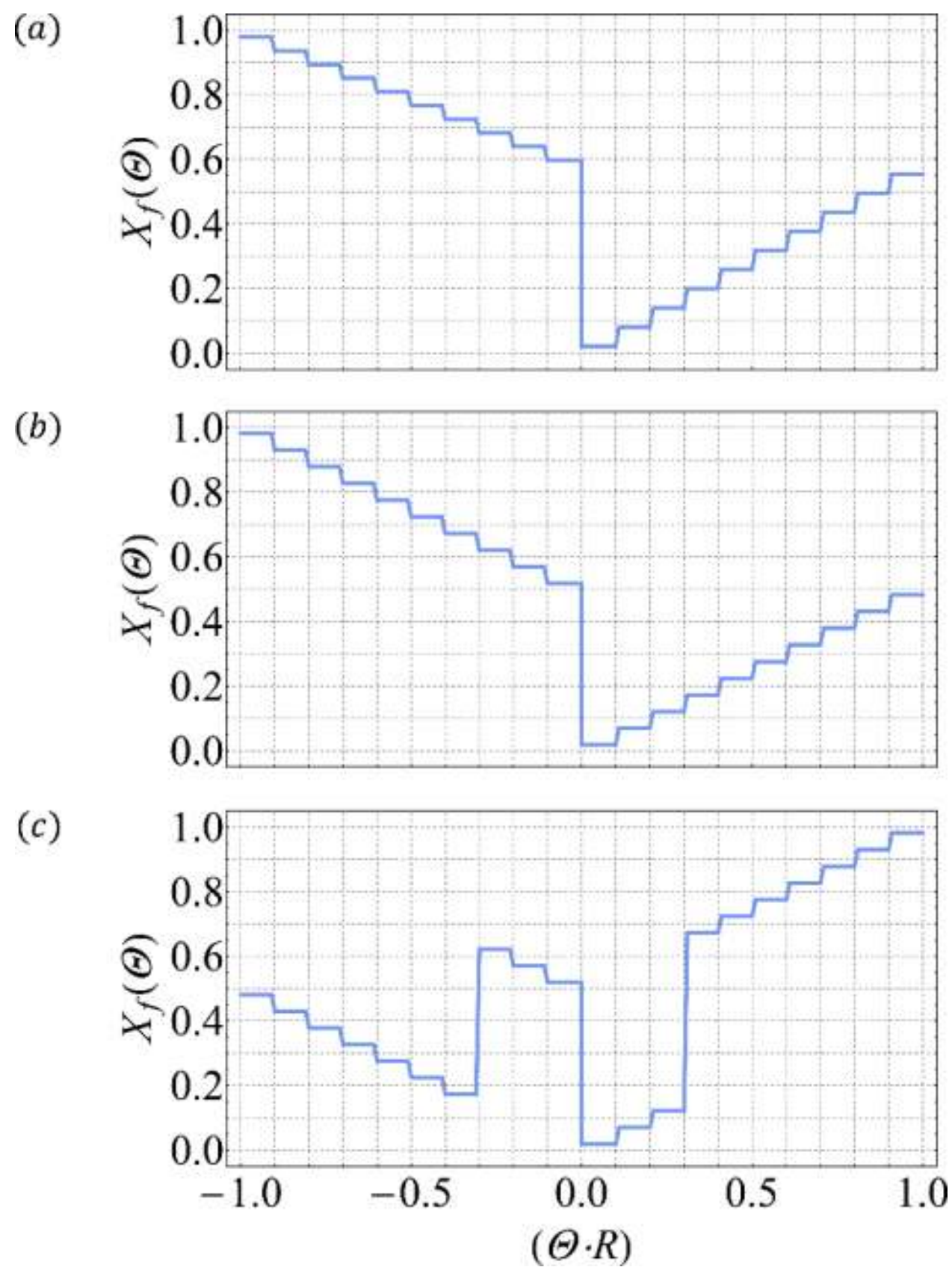

*FIG. A.4.* ***Plot of coordinate mapping $\Theta - X_f$ for three key configurations used in article.*** *(a) Upper serpentine. (b) Parallel serpentine. (c) Crossover serpentine.*

***$X_f - \Theta$ Coordinate Mapping, $\Theta(X_f)$***

We now turn to map our solid field deformation onto the fluidic pressure field. The pressure source terms in equation (A.13) are governed by $\partial \widetilde{M}_e(\Theta,T)/\partial T$ and $\partial \widetilde{N}_e(\Theta,T)/\partial T$ we thus need to map our resultants onto the fluid field to derive their respective values. Using the same methodology and definition for bladder index $j$, table index $k$, $\Theta$ and $R$, we construct the coordinate mapping table: in the $X_f$ column, we advance from one bladder to the next. Every $j$ indexed bladder owns two rows for its start and end position. Connected bladders are addressed first; The first connected bladder is assigned $X_{f,init}^{s=1} = 0$ and end value $X_{f,end}^{s=1} = \mathcal{L}_b$ at the respective $(R \cdot j)$ . Each consecutive connected bladder is assigned an initial-value $X_{f,init}^{s} = X_{f,end}^{s-1} + \mathcal{L}_c$ up to $X_{f,end}^{s} = X_{f,init}^{s} + \mathcal{L}_b$. In the case of separate interconnected bladder arrays, the $X_{f,init}^{s}$ of the first bladder in the new array has $X_{f,init}^{s} = X_{f,end}^{s-1} + \epsilon$, where $X_{f,end}^{s-1}$ is the $X_{f,end}$ of the last bladder in the previous connected array. We then continue as before for consecutive connected bladders. This process is repeated until we finish all separate interconnected arrays. Next we address the disconnected bladders by order of $\Omega_d$. Bladders are assigned an initial value $X_{f,init}^{q} = X_{f,end}^{q-1} + \epsilon$ and $X_{f,end}^{q} = X_{f,init}^{q} + \mathcal{L}_b$, where for the first disconnected bladder $X_{f,end}^{q-1}$ is the $X_{f,end}$ of the last bladder in the connected array. For the $(\Theta \cdot R)$ column we set

$$(\Theta \cdot R)_{2k-1} = (\Theta \cdot R)_{2k} = \begin{Bmatrix} k \le j\,, & \dfrac{2k-1}{n} \\ k > j\,, & -\left(\dfrac{2k-1}{n} - 1\right) \end{Bmatrix}, \tag{A.61}$$

representing the mid-point coordinate of respective bladder along $\Theta$ dimension such that along the entire length of bladder $(R \cdot j)$ we maintain our $\Theta$ coordinate. The mapping detailed above generates a $C^0$ continuous mapping function. Fig. A.5 illustrates the plot of coordinate mapping for three key examples, upper serpentine (panel a), parallel serpentine (panel b) and crossover (panel c). Tabulated data used in plots is presented in Table A.2a, A.2b and A.2c respectively in §A.6.

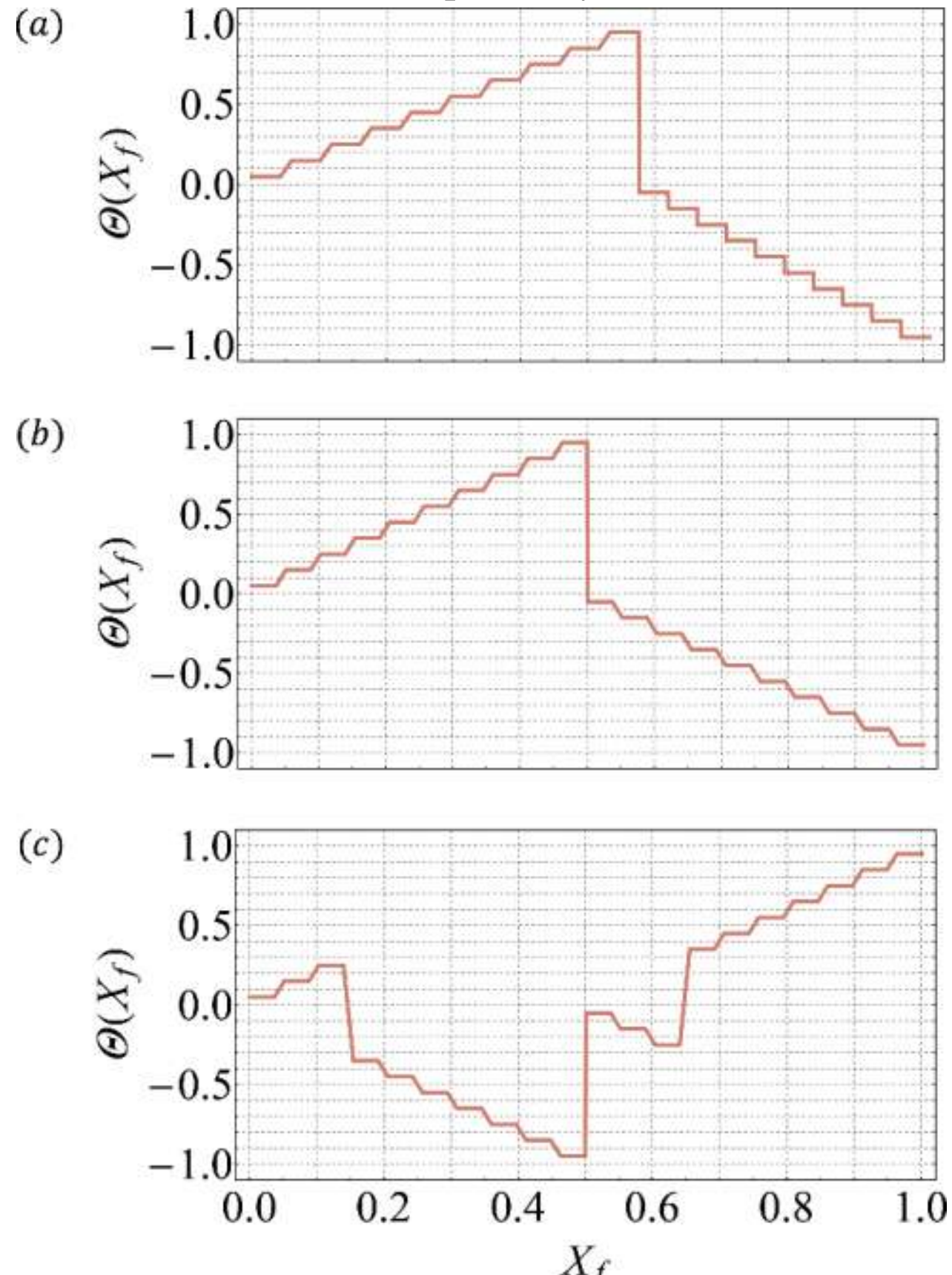

*FIG. A.5.* ***Plot of coordinate mapping*** $\mathbf{X_f - \Theta}$ ***for three key configurations used in article.*** *(a) Upper serpentine. (b) Parallel serpentine. (c) Crossover serpentine.*

### A.2.6. Formulation of $R(X_f)$

The function $R(X_f)$ is a function used to identify the position of a given bladder in upper or lower row. $R = 1$ indicated the upper bladder and $R = -1$ for lower bladder respectively. It is derived as a $C^0$ continuous function from the coordinate mapping $X_f - \Theta$ as

$$R(X_f) = \begin{Bmatrix} \Theta(X_f) = 0\,, & 0 \\ \Theta(X_f) \neq 0\,, & \dfrac{\Theta(X_f)}{|\Theta(X_f)|} \end{Bmatrix}. \tag{A.62}$$

### A.2.7. Formulation of $A_1(\widetilde{N}_e,\Theta)/\partial\widetilde{N}_e$ , $\partial A_1(\widetilde{M}_e,\Theta)/\partial\widetilde{M}_e$ , $\partial A_1(P,\Theta)/\partial P$ and $Q_1\left(A(X_f,P)\right)$

The functions $A_{N_1}(\widetilde{N}_e,\Theta)/\partial\widetilde{N}_e$ , $\partial A_{M_1}(\widetilde{M}_e,\Theta)/\partial\widetilde{M}_e$ , $\partial A_{p_1}(P,\Theta)/\partial P$ represent the change in cross section due to section internal resultants $\widetilde{N}_e$, $\widetilde{M}_e$ and pressure $P$. The function $Q_1\left(A(X_f,P)\right)$ stands for the fluidic cross section permeability. The above mentioned functions' value dependents on the fluidic cross section in question being that of a bladder or a connective tube. As such, their value is derived directly from the coordinate mapping $X_f-\Theta$ as a $C^0$ continuous function with $\Gamma(X_f)$ as the differentiating function

$$\Gamma(X_f)=\begin{Bmatrix}\dfrac{\partial\Theta(X_f)}{X_f}=0\,, & 1\\ \dfrac{\partial\Theta(X_f)}{X_f}\neq 0\,, & r_c\end{Bmatrix}. \tag{A.63}$$

We define $r_c$ as the ratio of quantitative value of respective measure in a tube divided by that of a bladder. We can now define

$$\frac{\partial A_{N_1}(\widetilde{N}_e,\Theta)}{\partial\widetilde{N}_e}=\frac{\partial A_{N_1}}{\partial\widetilde{N}_e}\,\Gamma(X_f)\,, \tag{A.64}$$

$$\frac{\partial A_{M_1}(\widetilde{M}_e,\Theta)}{\partial\widetilde{M}_e}=\frac{\partial A_{M_1}}{\partial\widetilde{M}_e}\,\Gamma(X_f)\,, \tag{A.65}$$

$$\frac{\partial A_{p_1}(P,\Theta)}{\partial P}=\frac{\partial A_{p_1}}{\partial P}\,\Gamma(X_f)\,. \tag{A.66}$$

For the case of (A.64) - (A.66) $r_c=0$ as connective tube segments do not change their cross section area due to force and moment resultants (being external to the structure) nor do they do due to pressure being constant cross section elastic tubes. As such it is of note to mention that for the connective tubes the fluidic governing equation (A.13) degenerates to $\partial^2 P/\partial {X_f}^2=0$, lending to the understanding of their contribution to fluid domain solution as a source for a steady state linear pressure gradient along their length.

Next we obtain $Q_1$ as a function of $X_f$ as we progress from bladder to tube and vice versa along the fluidic domain. Calculating the physical value of $q_1^c$ and $q_1^b$, see §A.4.2, we then set our bladder-tube scaling argument for $Q_1$ using (A.12) separately for a tube $Q_1^c=q_1^c/q_1^{c*}$ and bladder $Q_1^b=q_1^b/q_1^{b*}$ and define

$$Q_1\left(A(X_f,P)\right)=Q_1^b\Gamma(X_f)\,, \tag{A.67}$$

where $r_c=Q_1^c/Q_1^b$ such that we alternate between $Q_1^b$ or $Q_1^c$ respective to $X_f$ position.

### A.2.8. Setting Fluidic Field intervals, Boundary and Initial Conditions

We now have to divide our fluid field domain to segregated intervals. Each interconnected bladder-tube array is set as one $C^1$ continuous domain starting at the initial value of the first interconnected bladder in the array $X_{f,init}^s$, up to and including the initial value of the first bladder $X_{f,init}^s$ or $X_{f,init}^q$ at the successive domain. Disassociated bladders possess a $C^0$ discontinuity between them, such that there is no fluidic pressure propagation between disassociated bladders. The start and end coordinate of each interval is derived directly from the $X_f$ column of the $X_f-\Theta$ Coordinate Mapping. See Table A.2 at §A.6.

Once fluidic domain intervals are defined, boundary and initial conditions are set respective to system setup. An interval whose end i.e. whose bladder's end has a pressure inlet, is set with a Dirichlet condition

$$P(\beta,T) = \mathrm{K}_{11}(T)\,, \tag{A.68}$$

whereas for a sealed bladder, a Neumann condition is set

$$\left.\left(\frac{\partial P}{\partial X_f}\right)\right|_{(\beta,T)} = \mathrm{K}_{12}(T)\,, \tag{A.69}$$

where $\beta$ is $X_f$ coordinate at either end of the respective interval boundary. Last an initial conditions are set

$$P(X_f,0) = \mathrm{K}_{13}(X_f)\,. \tag{A.70}$$

## A.3. Model Analysis Summary

In this section, we provide a concise overview of the above proposed model. Solid field governing equations in non-dimensional form are given for the $X_s$ direction

$$\begin{aligned} X_s\colon \Pi_1\left(\frac{\partial^2 U_1}{\partial T^2}\right) = B_x &+ \left[\Pi_2\,\frac{\partial \tilde{N}_e}{\partial \Theta} - \Pi_3\tilde{\alpha}_e\tilde{V}_e\right]{}^sE_{tx} + \left[\Pi_4\,\frac{\partial \tilde{V}_e}{\partial \Theta} + \Pi_5\tilde{\alpha}_e\tilde{N}_e\right]{}^sE_{nx} \\ &+ \frac{\partial}{\partial \Theta}\left[\frac{1}{\tilde{d}_{33}}\left(\left(\Pi_6 B_x^1 - \Pi_7 Y^{11}\frac{\partial^2 \tilde{d}_{1x}}{\partial T^2}\right)\tilde{d}_{3x}\right.\right. \\ &\left.\left.+\left(\Pi_8 B_z^1 - \Pi_9 Y^{11}\frac{\partial^2 \tilde{d}_{1z}}{\partial T^2}\right)\tilde{d}_{3z}\right)\tilde{d}_{1x}\right] \end{aligned} \tag{A.44}$$

and in the $Z_s$ direction

$$\begin{aligned} Z_s\colon \Pi_1\left(\frac{\partial^2 U_3}{\partial T^2}\right) = B_z &+ \left[\Pi_2\,\frac{\partial \tilde{N}_e}{\partial \Theta} - \Pi_3\tilde{\alpha}_e\tilde{V}_e\right]{}^sE_{tz} + \left[\Pi_4\,\frac{\partial \tilde{V}_e}{\partial \Theta} + \Pi_5\tilde{\alpha}_e\tilde{N}_e\right]{}^sE_{nz} \\ &+ \frac{\partial}{\partial \Theta}\left[\frac{1}{\tilde{d}_{33}}\left(\left(\Pi_6 B_x^1 - \Pi_7 Y^{11}\frac{\partial^2 \tilde{d}_{1x}}{\partial T^2}\right)\tilde{d}_{3x}\right.\right. \\ &\left.\left.+\left(\Pi_8 B_z^1 - \Pi_9 Y^{11}\frac{\partial^2 \tilde{d}_{1z}}{\partial T^2}\right)\tilde{d}_{3z}\right)\tilde{d}_{1z}\right], \end{aligned} \tag{A.45}$$

where $\Pi_1 = \left(t_s^*/t_f^*\right)^2$, $\Pi_2 = N_e^*/(l_s m b^*)$, $\Pi_3 = (\alpha_e^* V_e^*)/(m b^*)$, $\Pi_4 = V_e^*/(l_s m b^*)$, $\Pi_5 = (\alpha_e^* N_e^*)/(m b^*)$, $\Pi_6 = \Pi_8 = (d_1^* b^{1*} d_3^*)/(l_s d_{33}^* b^*)$, $\Pi_7 = \Pi_9 = \left((d_1^* y^{11*} d_3^*)/(l_s^2\, d_{33}^*)\right)\left(t_s^*/t_f^*\right)^2$. These are supplemented by distributed traction force per unit mass and first moment per unit mass vectors respectively,

$$\boldsymbol{B} = \left(\frac{\rho_s l_s^2}{m} B_{b\,x} + B_{c\,x}\;,\;\; \frac{\rho_s l_s^2}{m}\left(\frac{b_z^*}{b_x^*}\right) B_{b\,z} + \left(\frac{b_z^*}{b_x^*}\right) B_{c\,z}\right), \tag{A.46}$$

$$\boldsymbol{B}^1 = \left(\frac{\rho_s l_s^2}{m} B_{b\,x}^1 + B_{c\,x}^1\,,\;\; \frac{\rho_s l_s^2}{m}\left(\frac{b_z^{1^*}}{b_x^{1^*}}\right) B_{b\,z}^1 + \left(\frac{b_z^{1^*}}{b_x^{1^*}}\right) B_{c\,z}^1\right). \tag{A.47}$$

The required constitutive laws are

$$\widetilde{N}_e = \left(\tilde{\lambda}_e - \frac{1}{\lambda_e^*}\right), \tag{A.34}$$

$$\widetilde{M}_e = \tilde{\alpha}_e\,, \tag{A.35}$$

$$\tilde{V}_e = -\frac{1}{{}^s\tilde{d}_{33}^{1/2}}\,\frac{\partial \widetilde{M}_e}{\partial \Theta}\,, \tag{A.36}$$

and the intrinsic kinematic variables

$$\tilde{\lambda}_e = \frac{1}{\lambda_e^*}\underbrace{\frac{{}^s\tilde{d}_{33}^{1/2}}{{}^s\widetilde{D}_{33}^{1/2}}}_{\tilde{\lambda}_s} - \frac{\lambda_p^*}{\lambda_e^*}\underbrace{\bar{P}(X_f)\frac{\partial \tilde{\lambda}_p}{\partial(\bar{p}/E)}}_{\tilde{\lambda}_p}, \tag{A.29}$$

$$\tilde{\alpha}_e = \frac{1}{l_s\alpha_e^*}\underbrace{\frac{\tilde{d}_{1x,3}\,\tilde{d}_{3x} + \tilde{d}_{1z,3}\,\tilde{d}_{3z}}{{}^s\tilde{d}_{33}^{1/2}\,{}^s\widetilde{D}_{33}^{1/2}}}_{\tilde{\alpha}_s} + \frac{\alpha_p^*}{\alpha_e^*}\underbrace{P'(X_f)\frac{\partial \tilde{\alpha}_p}{\partial(p'/E)}}_{\tilde{\alpha}_p}. \tag{A.30}$$

The pressure field produces a source term for curvature and extension per solid domain coordinate $\Theta$. Defining these terms requires us to map our fluidic domain solution using the $X_f(\Theta)$ mapping, see §A.2.5 such that we are able to define the effective pressure generating curvature and extension $P'(X_f) = \left(P_d(X_f) - P_u(X_f)\right)$ and $\bar{P}(X_f) = \left(P_d(X_f) + P_u(X_f)\right)$ respectively. Last we require six boundary conditions and four initial conditions for a well-posed problem, see in detail equations (A.49) - (A.58). The fluidic domain is governed by the non-dimensional equation

$$\begin{aligned} &-\left(\frac{\partial^2 P}{\partial {X_f}^2}\cdot Q_1\left(A\left(X_f, P\right)\right) + \frac{\partial P}{\partial X_f}\cdot\left(\frac{\partial Q_1}{\partial X_f}\right)\right) \\ &\quad + \left(\frac{\partial A_{p_1}(P,\Theta)}{\partial P}\frac{\partial P}{\partial T} + \mathrm{R}\left(X_f\right)\frac{\partial A_{M_1}\left(\widetilde{M}_\mathrm{e},\Theta\right)}{\partial \widetilde{M}_e}\frac{\partial \widetilde{M}_e(\Theta,T)}{\partial T}\right. \\ &\quad \left. + \left|\left(\mathrm{R}\left(X_f\right)\right)\right|\frac{\partial A_{N_1}\left(\widetilde{N}_\mathrm{e},\Theta\right)}{\partial \widetilde{N}_e}\frac{\partial \widetilde{N}_e(\Theta,T)}{\partial T}\right) = 0\,, \end{aligned} \tag{A.13}$$

Where $\partial A_1(P,\Theta)/\partial P$ , $\partial A_1\left(\widetilde{N}_\mathrm{e},\Theta\right)/\partial \widetilde{N}_e$ , $\partial A_1\left(\widetilde{M}_\mathrm{e},\Theta\right)/\partial \widetilde{M}_e$ represent the change in cross section due to pressure and section internal resultants $\widetilde{N}_e$, $\widetilde{M}_e$ respectively. We define

$$\frac{\partial A_{N_1}\left(\widetilde{N}_\mathrm{e},\Theta\right)}{\partial \widetilde{N}_e} = \frac{\partial A_{N_1}}{\partial \widetilde{N}_e}\,\Gamma\left(X_f\right), \tag{A.64}$$

$$\frac{\partial A_{M_1}\left(\widetilde{M}_\mathrm{e},\Theta\right)}{\partial \widetilde{M}_e} = \frac{\partial A_{M_1}}{\partial \widetilde{M}_e}\,\Gamma\left(X_f\right), \tag{A.65}$$

$$\frac{\partial A_{p_1}(P,\Theta)}{\partial P} = \frac{\partial A_{p_1}}{\partial P}\,\Gamma\left(X_f\right), \tag{A.66}$$

and a bladder-tube differentiating function $\Gamma(X_f)$

$$\Gamma\left(X_f\right) = \left\{\begin{matrix} \dfrac{\partial\Theta\left(X_f\right)}{X_f} = 0\,, & 1 \\ \dfrac{\partial\Theta\left(X_f\right)}{X_f} \neq 0\,, & r_c \end{matrix}\right\}. \tag{A.63}$$

We define $\mathcal{r}_c$ as the ratio of quantitative value of respective measure in a tube divided by that of a bladder. $\Theta(X_f)$ is the coordinate mapping function discussed in §A.2.5 mapping our resultants and change in cross section onto the fluid domain to derive their respective values. We then define

$$\frac{\partial A_{M_1}}{\partial \widetilde{M}_e} = \begin{Bmatrix} R = 1 \quad \cap \quad \widetilde{M}_e \geq 0\,, & \dfrac{\partial A_{M_1}}{\partial \widetilde{M}_e} = \left(\dfrac{\partial A_{M_1}}{\partial \widetilde{M}_e}\right)_- \\ R = 1 \quad \cap \quad \widetilde{M}_e < 0\,, & \dfrac{\partial A_{M_1}}{\partial \widetilde{M}_e} = \left(\dfrac{\partial A_{M_1}}{\partial \widetilde{M}_e}\right)_+ \\ R = -1 \quad \cap \quad \widetilde{M}_e \geq 0\,, & \dfrac{\partial A_{M_1}}{\partial \widetilde{M}_e} = \left(\dfrac{\partial A_{M_1}}{\partial \widetilde{M}_e}\right)_+ \\ R = -1 \quad \cap \quad \widetilde{M}_e < 0\,, & \dfrac{\partial A_{M_1}}{\partial \widetilde{M}_e} = \left(\dfrac{\partial A_{M_1}}{\partial \widetilde{M}_e}\right)_- \end{Bmatrix}, \tag{A.15}$$

and

$$\frac{\partial A_{N_1}}{\partial \widetilde{N}_e} = \begin{Bmatrix} \widetilde{N}_e \geq 0\,, & \dfrac{\partial A_{N_1}}{\partial \widetilde{N}_e} = \left(\dfrac{\partial A_{N_1}}{\partial \widetilde{N}_e}\right)_- \\ \widetilde{N}_e < 0\,, & \dfrac{\partial A_{N_1}}{\partial \widetilde{N}_e} = \left(\dfrac{\partial A_{N_1}}{\partial \widetilde{N}_e}\right)_+ \end{Bmatrix}, \tag{A.16}$$

where $\left(\partial A_{M_1}/\partial \widetilde{M}_e\right)_+$ and $\left(\partial A_{M_1}/\partial \widetilde{M}_e\right)_-$ represent the change in cross section due to section moment resultant, that will induce respectively a positive and negative pressure, and $\left(\partial A_{N_1}/\partial \widetilde{N}_e\right)_+$ , $\left(\partial A_{N_1}/\partial \widetilde{N}_e\right)_-$ as the change in cross section due to section normal force resultant, that will induce respectively a positive and negative pressure, these values are calculated per system, as detailed in §A.4.2. Last we set

$$Q_1\left(A(X_f, P)\right) = Q_1^b \Gamma(X_f)\,, \tag{A.67}$$

where and $\mathcal{r}_c = Q_1^c/Q_1^b$ such that we alternate between bladder value $Q_1^b$ and tube's $Q_1^c$ respective to $X_f$ position. Finishing our model, we implement fluidic field intervals, boundary and initial conditions as explained in detail in §A.2.8, and with that we have concluded the survey of our purposed model formulation.

## A.4. Model Validation

Validating the purposed model we set a full scale FEM 3D fluid-structure interaction model using COMSOL 5.3a enabling geometric nonlinearity yet limiting for a Hookean linear elastic material. In §A.4.1 we solve the 3D solid field alone, with fluid effects set negligible, to calculate the coefficients $f_e$ and $f_i$. In §A.4.2 we purpose a full scale 3D fluid-structure Interaction model (hereafter referred to as FSI) with all bladders at both upper and bottom rows disassociated, and use it to derive the coupling coefficients $\partial a_{p_1}/\partial p$, $\partial a_{M_1}/\partial M_e$, $\partial a_{N_1}/\partial N_e$, $\partial \psi/\partial(p'/E)$, $\partial \zeta/\partial(\bar{p}/E)$. In §A.4.3 we present the results for a full scale 3D FEM FSI model with tubes interconnecting the upper row bladders in a serpentine configuration with a Hagen-Poiseuille modeled flow for the connective tubes, and use it to validate results of an equivalent system solved using the above purposed model as summarized in §A.3.

In all subsections herein we use a solid beam with and internal honeycomb structure composed of bladders. The physical parameters of our system are $l_s = 0.1[m]$, $h_s = 0.0216[m]$, $w_s = 0.050[m]$ and $l_b = 0.046[m]$, $h_b = w_b = 0.0078[m]$, $E = 2[MPa]$, $\rho_s = 950[Kg/m^3]$. The honeycomb structure is composed of 20 bladders, 10 at the top 10 at the bottom.

### A.4.1. Derivation of Solid Field Coefficients

In this section we aim to calculate the solid domain coefficients $f_e$ and $f_i$. For this purpose we set up a test system may be either an experimental or an equivalent 3D FEM computational model. The test system must meet the following criteria: it must have the same external cross section geometry $h_s, w_s$ and it must house the same bladder geometry as defined by parameters $h_b, w_b, l_b$ and $\phi$ required to define the periodic unit of the honeycomb-like structure. We modify our proposed model, see summarized in §A.3, and decoupled fluid and solid domains by setting $\lambda_p^* = 0, \alpha_p^* = 0$, thus eliminating the fluid influence on the solid. In addition we set $\partial a_{M_1}/\partial M_e = 0$ and $\partial a_{N_1}/\partial N_e = 0$ such that we also eliminate the solid influence from the fluid. We thus effectively solve for the solid structure honeycomb alone eliminating the influence of fluid domain from the solid domain solution. Next we calculate the mass fraction of the honeycomb structure, $f_m = 0.481[1]$, and mass per unit length $m = 0.494[Kg/m]$, where we set $\rho_f = 1[Kg/m^3]$ as would be the case for any honeycomb structure with embedded cavities filled with air and open to atmospheric pressure. It is of note, that while there is no stiff criteria as to the number of bladder $n$; the larger the sample of bladders over which the parameters are averaged, the lesser the error will be due to localized phenomena such as, large strains localized disproportionately in a small region of the material.

In Fig A.6, A.7 we see the end result of the iterative process of calculating the structural effective stiffness coefficients $f_e$ and $f_i$. We begin with a geometrical initial guess at index $j = 0$ to be $f_i^{j=0} = I_{hc}/I_{full}$, where $I_{hc}$ and $I_{full}$ are section moment of inertia of the honeycomb and full section respectively, and $f_e^{j=0} = A_{hc}/A_{full}$, where $A_{hc}$ and $A_{full}$ are the section area of the honeycomb and full section respectively. We then iteratively modify $f_e^j$ and $f_i^j$ and recalculate $N_e^{*j}$, $V_e^{*j}$ and $(B_x, B_z)$ from (A.37), (A.39) and (A.46) respectively, until we match the model solution matches that of the 3D FEM. Once complete we have effectively deduced the structural stiffness coefficients for tension $f_e$ and bending $f_i$ taking into account stress concentration considerations as are prevalent in such holey structures.

In Fig. A.6 we illustrate the step response of a cantilever honeycomb-like beam to an extensional distributed tangential force in the z-axis direction $b_{c_z} = 3000[N/m] \cdot S(t)$ . The function $S(t) = L/(1 + exp(-\kappa(t - t_0)))$ is the sigmoid function, with max value with $L[1] = 1$, curve steepness $k[1/\text{sec}] = 68$ and midpoint location along t-axis $t_0[\text{sec}] = 1/f_{dim}$ *with* $f_{dim} = 16[Hz]$.

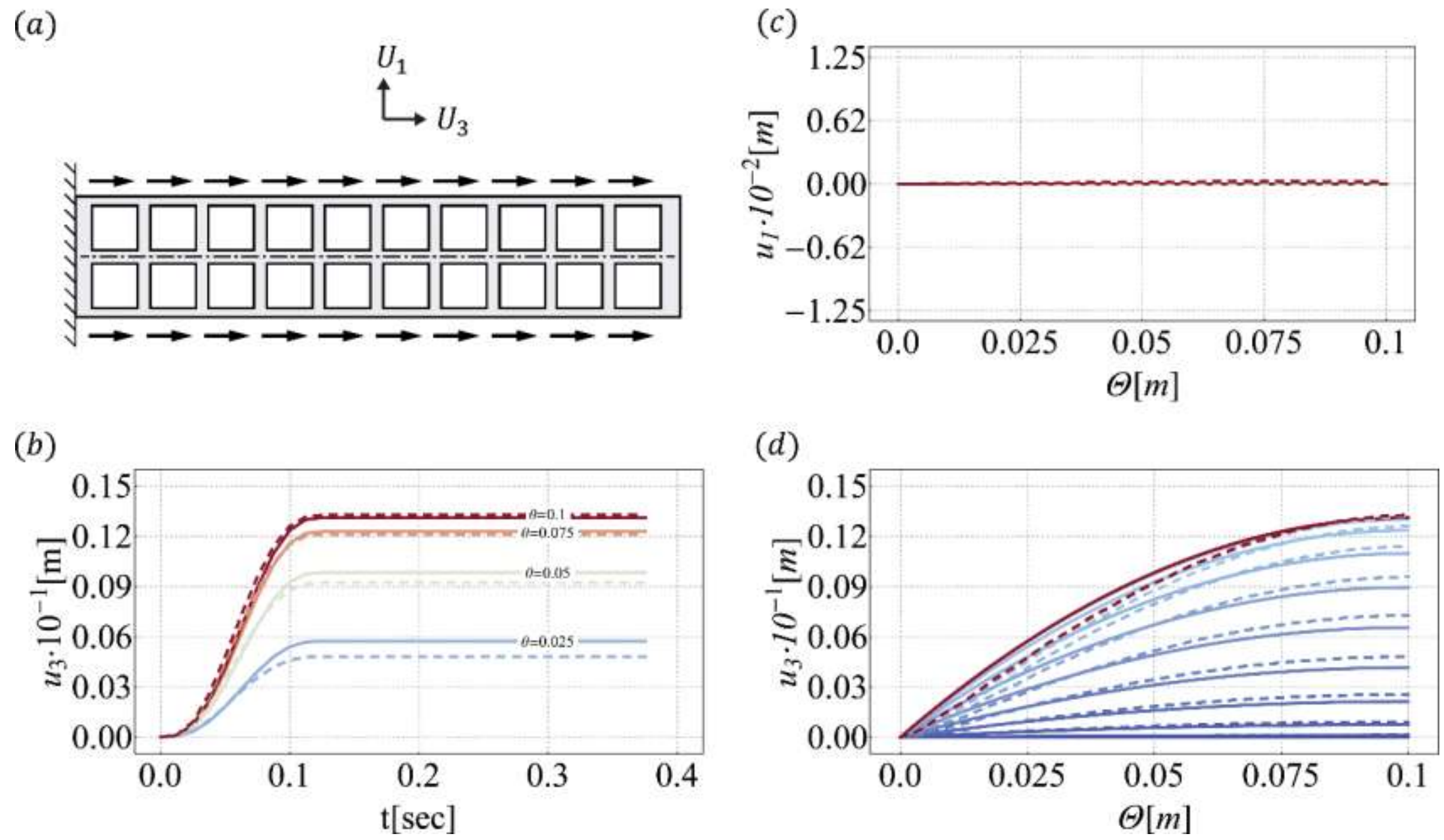

*FIG. A.6.* ***Illustration of a cantilever honeycomb-like structure used in the solid field validation.*** *Structure undergoes distributed tangential load* $b_{c_z} = 3000[N/m] \cdot S(t)$. *The function* $S(t) = L/(1 + \exp(-\kappa(t - t_0)))$ *is the sigmoid function, with max value* $L[1] = 1$, *curve steepness* $k[1/sec] = 68$ *and midpoint location along t-axis* $t_0[sec] = 1/f_{dim}$ *with* $f_{dim} = 16[Hz]$. *(a) Illustration of the cantilever beam with external force distribution and displacement axis directions. (b) Comparison of selected points' evolution. (c) Structure displacement in* $u_1$ *direction. (d) Structure displacement in* $u_3$ *direction. Model solution is presented by solid lines and full scale 3D FEM by dashed. Time evolution is presented via transition from blue to red color spectrum.*

In Fig. A.7 we illustrate the step response of a cantilever honeycomb-like beam to a distributed normal force in the x-axis direction $b_{c_x}(t) = 100 \cdot S(t)\ [N/m]$. The function $S(t) = L/(1 + exp(-\kappa(t - t_0)))$ is the sigmoid function, with max value $L[1] = 1$, curve steepness $k[1/\text{sec}] = 68$ and midpoint location along t-axis $t_0[\text{sec}] = 1/f_{dim}$ *with* $f_{dim} = 16[Hz]$.

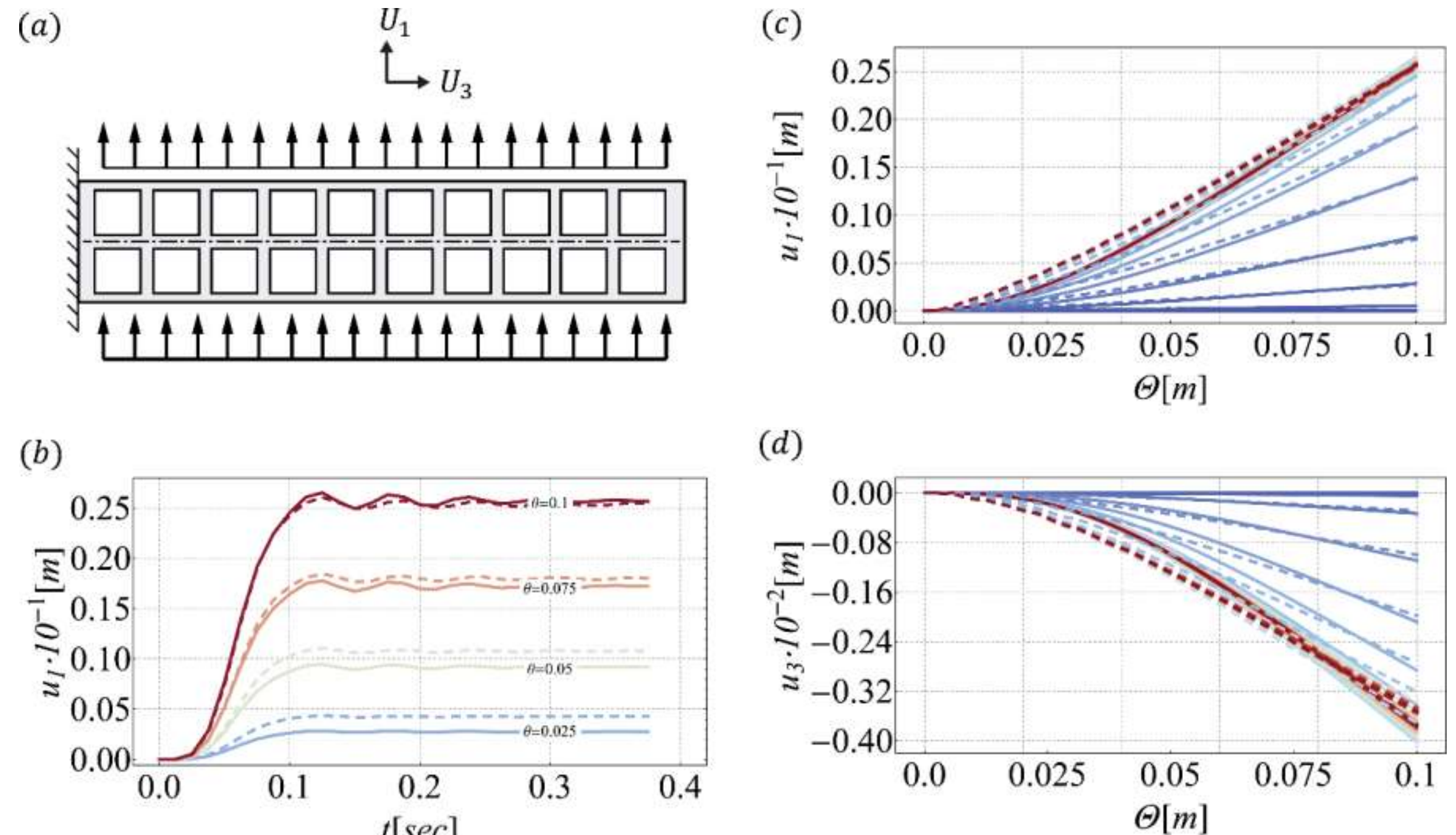

*FIG. A.7.* ***Illustration of a cantilever honeycomb-like structure used in the solid field validation.*** *Structure undergoing distributed normal load* $b_{c_x} = 100[\mathrm{N/m}] \cdot S(t)$. *The function* $S(t) = L/(1 + \exp(-\kappa(t - t_0)))$ *is the sigmoid function, with max value* $L[1] = 1$, *curve steepness* $k[1/\mathrm{sec}] = 68$ *and midpoint location along t-axis* $t_0[\mathrm{sec}] = 1/f_{\mathrm{dim}}$ *with* $f_{\mathrm{dim}} = 16[\mathrm{Hz}]$. *(a) Illustration of the cantilever beam with external force distribution and displacement axis directions. (b) Comparison of selected points' evolution. (c) Structure displacement in* $u_1$ *direction. (d) Structure displacement in* $u_3$ *direction. Model solution is presented by solid lines and full scale 3D FEM by dashed. Time evolution is presented via transition from blue to red color spectrum.*

In Conclusion, following the aforementioned iterative process, we find $f_e = 0.53$ and $f_i = 0.60$; good agreement can be seen for the solid field between the formulated model and the full-scale 3D FEM validation model. It is of note that while the 3D FEM validation model takes into account both the change in cross-section shape, area and tangential shear, our purposed model does not. It is this neglection in the formulated model that is responsible for the disagreement seen most evident at $\Theta < 0.4$ where the local cross section strains are of order $O(w_s h_s)$.

## A.4.2. Derivation of FSI Coupling Coefficients

In this section we expand on the process of deriving the required system parameters for the characterization of the fluid structure interaction coupling coefficients. To derive the coefficients we set up a test system following the same criteria as detailed in §A.4.1 in addition to having all bladders disassociated i.e. sealed at both ends along $l_b$, preventing flow and thus making the selection of fluid of no consequence as long as it remain incompressible in the given pressures and deformations of the process to follow. All input signals used in the various calculations are introduced using a sigmoid function $S(t) = L/(1 + exp(-\kappa(t - t_0)))$, with max value *of* $L[1] = 1$, curve steepness $k[1/\mathrm{sec}] = 68$ and midpoint location along t-axis $t_0[\mathrm{sec}] = 1/f_F$, *with* $f_F = 16[\mathrm{Hz}]$ being the excitation frequency for introduced pressures, normal forces and moments used, taken to be $f_F = o(t_f^*, t_s^*)$ thus ensuring a quasi-static evolution. In Fig. A.8 we present the result of numerical experiments used to calculate SLC parameters.

In Fig. A.8a, we present the calculation of $\partial a_{p_1}/\partial p$, the dimensional change in cross section area due to the fluid pressure. We quasi-statically introduce a growing, equally opposite negative and positive pressure to the upper and lower bladder rows respectively and evenly distributed within a row. We then measuring the fluid volume entering each bladder for the upper and lower rows separately. Averaging over the number of bladders per row $n/2$ and dividing by bladder length scale $l_b$, we thus obtain the averaged change in cross section due to fluid pressure per bladder for positive and negative pressure respectively. Fig. A.8a shows the proportional nature pressure-to-cross section change per bladder as observed over the range of $p = o(E)$ for both positive and negative pressure. Calculated value $\partial a_{p_1}/\partial p = 1.75 \cdot 10^{-10}[m^2/Pa]$.

In Fig. A.8b, we present the calculation of $\partial\zeta/\partial(\bar{p}/E)$, the dimensional change in length per cross section per unit of normalized pressure sum. We quasi-statically introduce a growing positive pressure into our bladders equally. Dividing the measured change in length of the test system by the number of bladders per row i.e. $n/2$ we thus obtain $\partial\zeta/\partial(\bar{p}/E)$. This parameter value maybe extrapolated over the negative pressure drawing on the conclusion from Fig. A.8a. Calculated value $\partial\zeta/\partial(\bar{p}/E) = 0.01\ [m]$.

In Fig. A.8c, we present the calculation of $\partial\psi/\partial(p'/E)$, the change in beam slope per cross sectionper unit normalized pressure difference. We quasi-statically introduce a growing, equally opposite negative and positive pressure respectively, to our top and bottom bladder rows simultaneously and evenly distributed within a row. We then measure the angle of the beam at tip and divide by the number of bladders in a row i.e. $n/2$. Calculated value $\partial\psi/\partial(p'/E) = 0.00381467/|P'|\ [rad]$.

In Fig. A.8d, we present the calculation of $\partial a_{M_1}/\partial M_e$, the change in fluidic cross section area due to solid domain moment resultant. We quasi-statically introduce a growing positive moment $M$ acting at beam end. We measure the cross section area of each bladder at $l_b/2$ separately for the upper or lower rows. We then average over the number of bladder per row $n/2$ to obtain $(\partial p/\partial M)_+$ and $(\partial p/\partial M)_-$ respectively. Equilibrium considerations of said load in a quasi-static system dictate $\partial a_{M_1}/\partial M = \partial a_{M_1}/\partial M_e$ respectively. Alternatively, one may measure the change in pressure per bladder due to moment $\partial p/\partial M$ for the upper and lower rows, and as in the direct method, average over the number of bladder per row $n/2$ separately to deduct $(\partial p/\partial M)_+$ and $(\partial p/\partial M)_-$ respectively, and then multiply by $\partial a_{p_1}/\partial p$ and we obtain $\partial a_{M_1}/\partial M$ . Calculated values $\left(\partial a_{M_1}/\partial M_e\right)_+ = 6.54 \cdot 10^{-6}[m^2/(N \cdot m)]$ and $\left(\partial a_{M_1}/\partial M_e\right)_- = 3.75 \cdot 10^{-6}[m^2/(N \cdot m)]$.

In Fig A.8e, we present the calculation of $\partial a_{N_1}/\partial N_e$, the change in fluidic cross section area due to solid domain normal force resultant. We again quasi-statically introduce a growing positive normal force $N$ acting at beam end. We measure the cross section area of each bladder at $l_b/2$, and then average over the total number of bladders $n$ to obtain $(\partial p/\partial N)_-$. Equilibrium considerations of said load in a quasi-static system dictate $\partial a_{N_1}/\partial N = \partial a_{N_1}/\partial N_e$ respectively. Alternatively, one may measure the change in pressure per bladder due to normal force $\partial p/\partial N$, and as in the direct method, average over the total number of bladders $n$ to deduct $(\partial p/\partial N)_-$. Then by multiplying by $\partial a_1/\partial p$ we obtain $\partial a_{N_1}/\partial N$. This parameter value maybe extrapolated over the positive pressure drawing on the conclusion from Fig. A.8d. Calculated values $\left(\partial a_{N_1}/\partial N_e\right)_- = 3.64 \cdot 10^{-8}[m^2/N]$ and $\left(\partial a_{N_1}/\partial N_e\right)_+ = 6.31 \cdot 10^{-8}[m^2/N]$.

Last, we calculate $q_1$. Reverting the Poisson equation (A.8) to its dimensional form, we obtain the flow rate $q = -((1/\mu)\partial p/\partial x_f)q_1^i$ where $q_1^i$ $(i = c, b)$. Applying order-or-magnitude analysis we obtain $q_1^{i*} = \tilde{C}^i\ r_{eff}^4$, where $r_{eff}$ is the effective flow path scale and $\tilde{C}^i \sim 4\pi$ is a dimensionless constant related to the configuration of the flow-path. We then separately solve for two domains: one for the connective tubes and another for the bladders. Setting $((1/\mu)\partial p/\partial x_f) = -1$ and no-slip condition set at the walls. We thus obtain respective $q_1^c$ and $q_1^b$ from $q$ solution, and deduce $Q_1^c = q_1^c/\tilde{C}^c r_{eff}^4$ and $Q_1^b = q_1^b/\tilde{C}^b r_{eff}^4$. Calculated values $q_1^b = 1.55 \cdot 10^{-12}[m^4]$ and $q_1^c = 6.72 \cdot 10^{-13}[m^4]$.

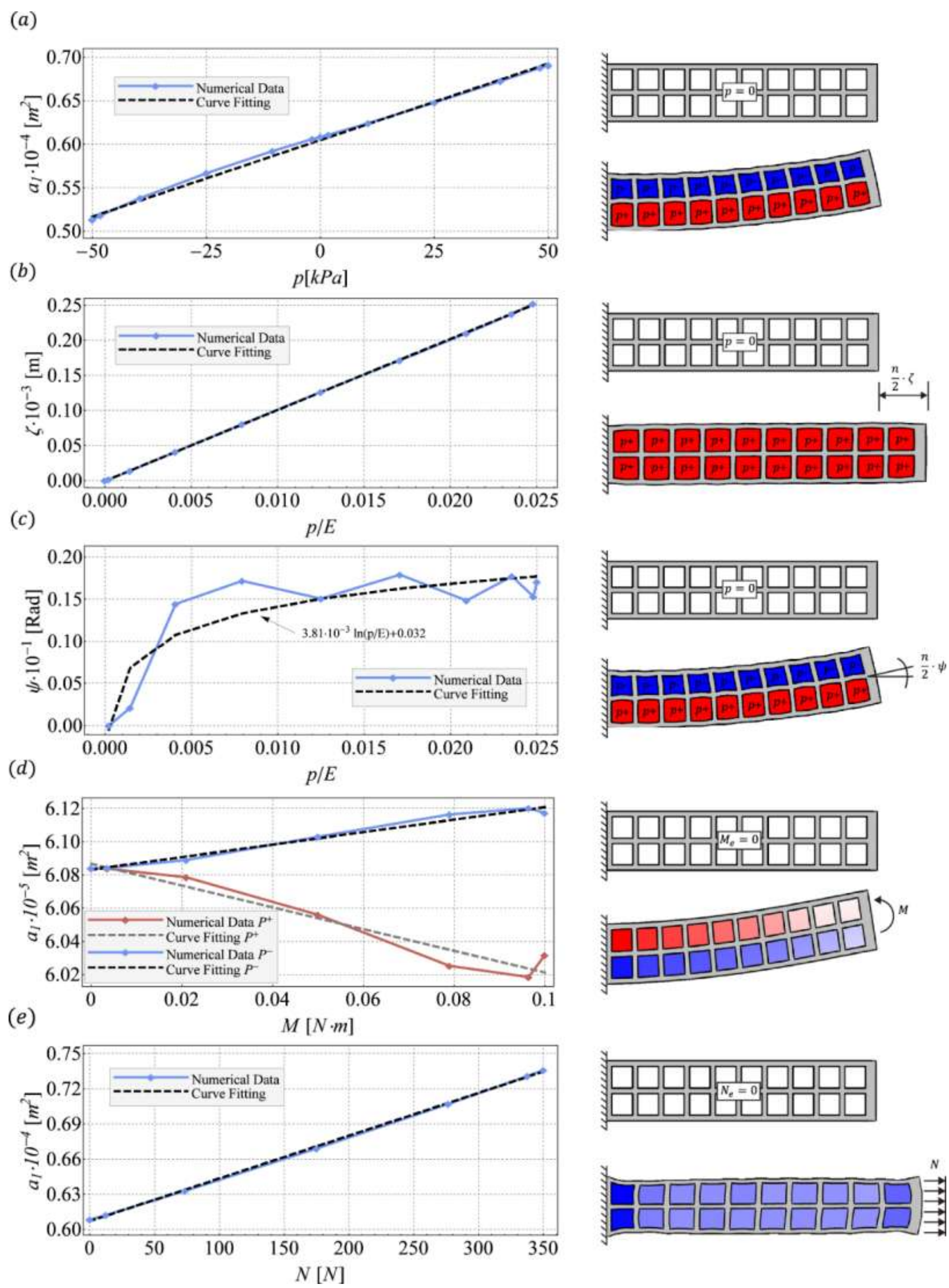

*FIG. A.8.* ***Left panels, result plot used to deduce coupling coefficients.*** *Right panels, Illustration of respective honeycomb-like structure used to calculate coefficient, where both relaxed state (top) and post actuation (bottom) illustrations are presented. High pressure is denoted in red spectrum and low pressure in blue for qualitative visual purposes. (a) Change in bladder cross section* $\mathrm{a_1[m^2]}$ *to pressure* $\mathrm{p[kPa]}$ *from which* $\partial \mathrm{a_{p_1}} / \partial \mathrm{p}$ *is derived. (b) Change in beam extension* $\zeta\mathrm{[m]}$ *to normalized pressure* $\mathrm{p/E\ [1]}$ *from which* $\partial\zeta / \partial(\mathrm{\bar{p}/E})$ *is derived. (c) Change in beam slope at* $\psi\mathrm{[rad]}$ *to normalize pressure* $\mathrm{p/E[1]}$ *from which to* $\partial\psi / \partial(\mathrm{p'/E})$ *is derived, (d) Change in bladder cross section* $\mathrm{a_1[m^2]}$ *to moment applied at beam tip* $\mathrm{M[N \cdot m]}$ *from which* $\partial \mathrm{a_{M_1}} / \partial \mathrm{M_e}$ *is derived. (e) Change in bladder cross section* $\mathrm{a_1[m^2]}$ *to normal force applied* $\mathrm{N[N]}$ *at beam tip from which* $\partial \mathrm{a_{N_1}} / \partial \mathrm{N_e}$ *is derived. Solid lines represent linear interpolation of numerical data points noted by diamond markers. Dashed lines represent least square method curve fitting.*

In conclusion, using the reduced geometry system above we were able to calculate the SLC coupling parameters as are used to characterize our system's two-way-coupled fluid structure interaction where, $\partial a_{p_1}/\partial p = 1.75 \cdot 10^{-10}[m^2/Pa]$ , $\partial\zeta/\partial(\bar{p}/E) = 0.01\ [m]$ , $\partial\psi/\partial(p'/E) = 0.00381467/|P'|\ [rad]$ , $\left(\partial a_{M_1}/\partial M_e\right)_+ = 6.54 \cdot 10^{-6}[m^2/(N \cdot m)]$ , $\left(\partial a_{M_1}/\partial M_e\right)_- = 3.75 \cdot 10^{-6}[m^2/(N \cdot m)]$ , $\left(\partial a_{N_1}/\partial N_e\right)_+ = 6.31 \cdot 10^{-8}[m^2/N]$ , $\left(\partial a_{N_1}/\partial N_e\right)_- = 3.64 \cdot 10^{-8}[m^2/N]$ . Last we calculate $q_1^b = 1.55 \cdot 10^{-12}[m^4]$ and $q_1^c = 6.72 \cdot 10^{-13}[m^4]$.

### A.4.3. Two Way Coupled FSI Validation

In this section we aim to examine the validity of purposed model in §A.3, and present a comparison to a 3D FEM solution of a fluid structure interaction model using COMSOL 5.3a commercial code. We set our system geometry as detailed above, see §A.4 opening statement. Three separate systems are set: Fig. A.9I pull under distributed tangential force in the z-axis $b_{c_z}(t) = 3000[N/m] \cdot S(t)$, Fig A.9II bend under a distributed normal force in the x-axis $b_{c_x}(t) = 100 \cdot S(t)\ [N/m]$ and Fig. A.9III cantilever beam with root oscillation $U_1(0,T) = U_{1,in}\,sin(2\pi\mathcal{F}_u \cdot T)$ *where* $F_u = t_f^* \cdot f_{dim} \approx 0.036$, *with* $f_{dim} = 8[Hz]$ *and* $U_{1,in} = 0.02$. The function $\mathrm{S(t) = L/(1 + \exp(-\kappa(t - t_0)))}$ is the sigmoid function, with max value $L[1] = 1$ , curve steepness $k[1/\mathrm{sec}] = 68$ and midpoint location along t-axis $t_0[\mathrm{sec}] = 1/f_{dim}$ *with* $f_{dim} = 16[Hz]$

Systems I and II use Glycerol 85% solution $\rho_f \approx 1000[Kg/m^3]$, $\mu_f = 0.1[Pa \cdot sec]$, system III uses Glycerol 55% solution $\rho_f \approx 1000[Kg/m^3]$, $\mu_f = 0.01[Pa \cdot sec]$, with reasoning being to best present solid and fluid field evolution over time. We calculate $f_m \approx 1[1]$ and $m \approx 1[Kg/m]$. All three systems have their top row bladders interconnected in a serpentine tubing configuration and lower row bladders disassociated. All piping needle valves are set open, so as to not disrupt the connective tube flow. Coupling parameters used in model are as calculated in §A.4. System setup is illustrated in Fig. A.9aI-III.

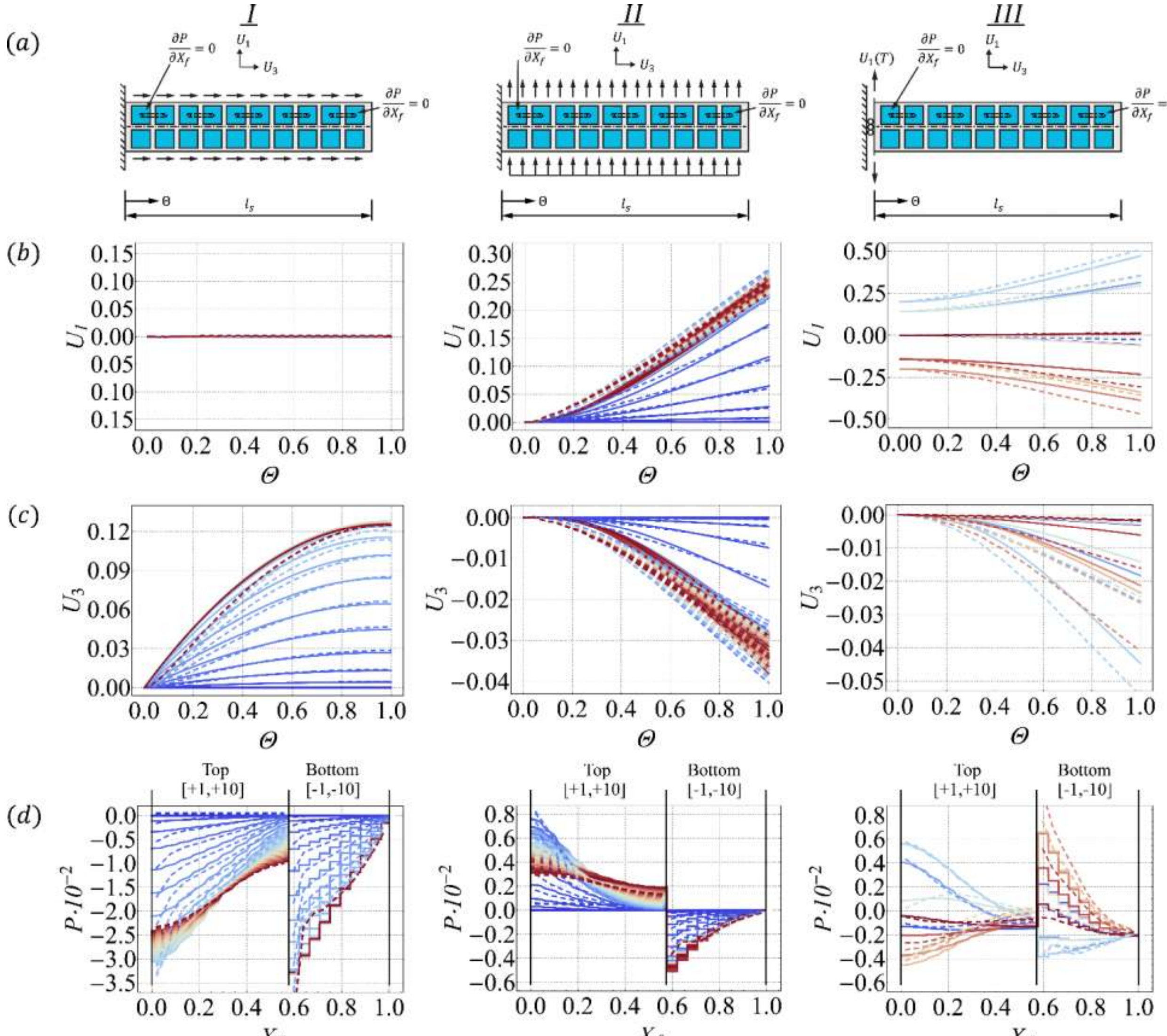

*FIG. A.9.* ***Illustration of system setup and results for three examined validation cases of upper serpentine configuration.*** *(a) System setup illustration. (b)* $U_1$ *deflection over length* $\Theta$. *(c)* $U_3$ *extension over length* $\Theta$. *(d)* P *over length* $X_f$. *Column I, depicts results for a cantilever beam under distributed tangential load in the z-axis* $b_{c_z}(t) = 3000[N/m] \cdot S(t)$, *Column II , cantilever beam under a distributed normal load in the x-axis* $b_{c_x}(t) = 100 \cdot S(t)\ [N/m]$ *and column III cantilever beam with root oscillation* $U_1(0,T) = U_{1,in} \sin(2\pi\mathcal{F}_u \cdot T)$ *where* $F_u = t_f^* \cdot f_{dim} \approx 0.036$, *with* $f_{dim} = 8[Hz]$ *and* $U_{1,in} = 0.02$. *The function* $S(t) = L/(1 + \exp(-\kappa(t - t_0)))$ *is the sigmoid function, with max value* $L[1] = 1$, *curve steepness* $k[1/sec] = 68$ *and midpoint location along t-axis* $t_0[sec] = 1/f_{dim}$ *with* $f_{dim} = 16[Hz]$. *Systems I and II use Glycerol 85% solution* $\rho_f \approx 1000[Kg/m^3]$, $\mu_f = 0.1[Pa \cdot sec]$ *as fluid whereas system III uses Glycerol 55% solution* $\rho_f \approx 1000[Kg/m^3]$, $\mu_f = 0.01[Pa \cdot sec]$. *Bladder* $(R \cdot j)$ *index range along* $X_f$ *coordinate are noted in square brackets in row (d). Model solution is presented by solid lines and full scale 3D FEM by dashed. Time evolution is presented via transition from blue to red color spectrum. Above settings are selected to best present solid and fluid field evolution over time.*

Good agreement is observed for both deflection and pressure fields for all three case. The under-prediction observed for $U_1$ in Fig. A.9aIII at both start and end cycle (bright red and blue lines) is kinematically matching the respective $U_3$ in Fig. A.9bIII. A possible source for this this under prediction of the model stems from the violation of the small local strains assumption at beam root, and matches the under prediction of pressure as seen predominantly at the disassociated lower row bladders, see Fig. A.9dIII at $X_f \approx 0.6$.

## A.5. Tabulated Data for $\Theta \rightarrow X_f$ Coordinate Mapping

*TABLE. A.1.* ***Tabulated data for $\Theta \rightarrow X_f$ coordinate mapping used in figure A.4.*** *(a) Upper serpentine. (b) Parallel serpentine. (c) Crossover.*

(a)

| $\Theta \cdot R$ | $X_f(\Theta)$ | $j \cdot R$ |
|---|---|---|
| −1. | 0.978814 | −10. |
| −0.91 | 0.978814 | −10. |
| −0.9 | 0.936441 | −9. |
| −0.81 | 0.936441 | −9. |
| −0.8 | 0.894068 | −8. |
| −0.71 | 0.894068 | −8. |
| −0.7 | 0.851695 | −7. |
| −0.61 | 0.851695 | −7. |
| −0.6 | 0.809322 | −6. |
| −0.51 | 0.809322 | −6. |
| −0.5 | 0.766949 | −5. |
| −0.41 | 0.766949 | −5. |
| −0.4 | 0.724576 | −4. |
| −0.31 | 0.724576 | −4. |
| −0.3 | 0.682203 | −3. |
| −0.21 | 0.682203 | −3. |
| −0.2 | 0.639831 | −2. |
| −0.11 | 0.639831 | −2. |
| −0.1 | 0.597458 | −1. |
| −0.001 | 0.597458 | −1. |
| 0. | 0.0211864 | 1. |
| 0.1 | 0.0211864 | 1. |
| 0.11 | 0.0805085 | 2. |
| 0.2 | 0.0805085 | 2. |
| 0.21 | 0.139831 | 3. |
| 0.3 | 0.139831 | 3. |
| 0.31 | 0.199153 | 4. |
| 0.4 | 0.199153 | 4. |
| 0.41 | 0.258475 | 5. |
| 0.5 | 0.258475 | 5. |
| 0.51 | 0.317797 | 6. |
| 0.6 | 0.317797 | 6. |
| 0.61 | 0.377119 | 7. |
| 0.7 | 0.377119 | 7. |
| 0.71 | 0.436441 | 8. |
| 0.8 | 0.436441 | 8. |
| 0.81 | 0.495763 | 9. |
| 0.9 | 0.495763 | 9. |
| 0.91 | 0.555085 | 10. |
| 1. | 0.555085 | 10. |

(b)

| $\Theta \cdot R$ | $X_f(\Theta)$ | $j \cdot R$ |
|---|---|---|
| −1. | 0.981618 | −10. |
| −0.91 | 0.981618 | −10. |
| −0.9 | 0.930147 | −9. |
| −0.81 | 0.930147 | −9. |
| −0.8 | 0.878676 | −8. |
| −0.71 | 0.878676 | −8. |
| −0.7 | 0.827206 | −7. |
| −0.61 | 0.827206 | −7. |
| −0.6 | 0.775735 | −6. |
| −0.51 | 0.775735 | −6. |
| −0.5 | 0.724265 | −5. |
| −0.41 | 0.724265 | −5. |
| −0.4 | 0.672794 | −4. |
| −0.31 | 0.672794 | −4. |
| −0.3 | 0.621324 | −3. |
| −0.21 | 0.621324 | −3. |
| −0.2 | 0.569853 | −2. |
| −0.11 | 0.569853 | −2. |
| −0.1 | 0.518382 | −1. |
| −0.001 | 0.518382 | −1. |
| 0. | 0.0183824 | 1. |
| 0.1 | 0.0183824 | 1. |
| 0.11 | 0.0698529 | 2. |
| 0.2 | 0.0698529 | 2. |
| 0.21 | 0.121324 | 3. |
| 0.3 | 0.121324 | 3. |
| 0.31 | 0.172794 | 4. |
| 0.4 | 0.172794 | 4. |
| 0.41 | 0.224265 | 5. |
| 0.5 | 0.224265 | 5. |
| 0.51 | 0.275735 | 6. |
| 0.6 | 0.275735 | 6. |
| 0.61 | 0.327206 | 7. |
| 0.7 | 0.327206 | 7. |
| 0.71 | 0.378676 | 8. |
| 0.8 | 0.378676 | 8. |
| 0.81 | 0.430147 | 9. |
| 0.9 | 0.430147 | 9. |
| 0.91 | 0.481618 | 10. |
| 1. | 0.481618 | 10. |

(c)

| $\Theta \cdot R$ | $X_f(\Theta)$ | $j \cdot R$ |
|---|---|---|
| −1. | 0.481618 | −10. |
| −0.91 | 0.481618 | −10. |
| −0.9 | 0.430147 | −9. |
| −0.81 | 0.430147 | −9. |
| −0.8 | 0.378676 | −8. |
| −0.71 | 0.378676 | −8. |
| −0.7 | 0.327206 | −7. |
| −0.61 | 0.327206 | −7. |
| −0.6 | 0.275735 | −6. |
| −0.51 | 0.275735 | −6. |
| −0.5 | 0.224265 | −5. |
| −0.41 | 0.224265 | −5. |
| −0.4 | 0.172794 | −4. |
| −0.31 | 0.172794 | −4. |
| −0.3 | 0.621324 | −3. |
| −0.21 | 0.621324 | −3. |
| −0.2 | 0.569853 | −2. |
| −0.11 | 0.569853 | −2. |
| −0.1 | 0.518382 | −1. |
| −0.001 | 0.518382 | −1. |
| 0. | 0.0183824 | 1. |
| 0.1 | 0.0183824 | 1. |
| 0.11 | 0.0698529 | 2. |
| 0.2 | 0.0698529 | 2. |
| 0.21 | 0.121324 | 3. |
| 0.3 | 0.121324 | 3. |
| 0.31 | 0.672794 | 4. |
| 0.4 | 0.672794 | 4. |
| 0.41 | 0.724265 | 5. |
| 0.5 | 0.724265 | 5. |
| 0.51 | 0.775735 | 6. |
| 0.6 | 0.775735 | 6. |
| 0.61 | 0.827206 | 7. |
| 0.7 | 0.827206 | 7. |
| 0.71 | 0.878676 | 8. |
| 0.8 | 0.878676 | 8. |
| 0.81 | 0.930147 | 9. |
| 0.9 | 0.930147 | 9. |
| 0.91 | 0.981618 | 10. |
| 1. | 0.981618 | 10. |

## A.6. Tabulated Data for $X_f \rightarrow \Theta$ Coordinate Mapping

*TABLE. A.2.* ***Tabulated data for*** $\mathbf{X_f} \rightarrow \boldsymbol{\Theta}$ ***coordinate mapping used in figure A.5****. (a) Upper serpentine. (b) Parallel serpentine. (c) Crossover.*

$(a)$

| $X_f$ | $\Theta(X_f)$ | $j \cdot R$ |
|---|---|---|
| 0. | 0.05 | 1. |
| 0.0423729 | 0.05 | 1. |
| 0.059322 | 0.15 | 2. |
| 0.101695 | 0.15 | 2. |
| 0.118644 | 0.25 | 3. |
| 0.161017 | 0.25 | 3. |
| 0.177966 | 0.35 | 4. |
| 0.220339 | 0.35 | 4. |
| 0.237288 | 0.45 | 5. |
| 0.279661 | 0.45 | 5. |
| 0.29661 | 0.55 | 6. |
| 0.338983 | 0.55 | 6. |
| 0.355932 | 0.65 | 7. |
| 0.398305 | 0.65 | 7. |
| 0.415254 | 0.75 | 8. |
| 0.457627 | 0.75 | 8. |
| 0.474576 | 0.85 | 9. |
| 0.516949 | 0.85 | 9. |
| 0.533898 | 0.95 | 10. |
| 0.576271 | 0.95 | 10. |
| 0.577271 | −0.05 | −1. |
| 0.619644 | −0.05 | −1. |
| 0.620644 | −0.15 | −2. |
| 0.663017 | −0.15 | −2. |
| 0.664017 | −0.25 | −3. |
| 0.70639 | −0.25 | −3. |
| 0.70739 | −0.35 | −4. |
| 0.749763 | −0.35 | −4. |
| 0.750763 | −0.45 | −5. |
| 0.793136 | −0.45 | −5. |
| 0.794136 | −0.55 | −6. |
| 0.836508 | −0.55 | −6. |
| 0.837508 | −0.65 | −7. |
| 0.879881 | −0.65 | −7. |
| 0.880881 | −0.75 | −8. |
| 0.923254 | −0.75 | −8. |
| 0.924254 | −0.85 | −9. |
| 0.966627 | −0.85 | −9. |
| 0.967627 | −0.95 | −10. |
| 1.01 | −0.95 | −10. |

$(b)$

| $X_f$ | $\Theta(X_f)$ | $j \cdot R$ |
|---|---|---|
| 0. | 0.05 | 1. |
| 0.0367647 | 0.05 | 1. |
| 0.0514706 | 0.15 | 2. |
| 0.0882353 | 0.15 | 2. |
| 0.102941 | 0.25 | 3. |
| 0.139706 | 0.25 | 3. |
| 0.154412 | 0.35 | 4. |
| 0.191176 | 0.35 | 4. |
| 0.205882 | 0.45 | 5. |
| 0.242647 | 0.45 | 5. |
| 0.257353 | 0.55 | 6. |
| 0.294118 | 0.55 | 6. |
| 0.308824 | 0.65 | 7. |
| 0.345588 | 0.65 | 7. |
| 0.360294 | 0.75 | 8. |
| 0.397059 | 0.75 | 8. |
| 0.411765 | 0.85 | 9. |
| 0.448529 | 0.85 | 9. |
| 0.463235 | 0.95 | 10. |
| 0.5 | 0.95 | 10. |
| 0.501 | −0.05 | −1. |
| 0.537765 | −0.05 | −1. |
| 0.552471 | −0.15 | −2. |
| 0.589235 | −0.15 | −2. |
| 0.603941 | −0.25 | −3. |
| 0.640706 | −0.25 | −3. |
| 0.655412 | −0.35 | −4. |
| 0.692176 | −0.35 | −4. |
| 0.706882 | −0.45 | −5. |
| 0.743647 | −0.45 | −5. |
| 0.758353 | −0.55 | −6. |
| 0.795118 | −0.55 | −6. |
| 0.809824 | −0.65 | −7. |
| 0.846588 | −0.65 | −7. |
| 0.861294 | −0.75 | −8. |
| 0.898059 | −0.75 | −8. |
| 0.912765 | −0.85 | −9. |
| 0.949529 | −0.85 | −9. |
| 0.964235 | −0.95 | −10. |
| 1.001 | −0.95 | −10. |

$(c)$

| $X_f$ | $\Theta(X_f)$ | $j \cdot R$ |
|---|---|---|
| 0. | 0.05 | 1. |
| 0.0367647 | 0.05 | 1. |
| 0.0514706 | 0.15 | 2. |
| 0.0882353 | 0.15 | 2. |
| 0.102941 | 0.25 | 3. |
| 0.139706 | 0.25 | 3. |
| 0.655412 | 0.35 | 4. |
| 0.692176 | 0.35 | 4. |
| 0.706882 | 0.45 | 5. |
| 0.743647 | 0.45 | 5. |
| 0.758353 | 0.55 | 6. |
| 0.795118 | 0.55 | 6. |
| 0.809824 | 0.65 | 7. |
| 0.846588 | 0.65 | 7. |
| 0.861294 | 0.75 | 8. |
| 0.898059 | 0.75 | 8. |
| 0.912765 | 0.85 | 9. |
| 0.949529 | 0.85 | 9. |
| 0.964235 | 0.95 | 10. |
| 1.001 | 0.95 | 10. |
| 0.501 | −0.05 | −1. |
| 0.537765 | −0.05 | −1. |
| 0.552471 | −0.15 | −2. |
| 0.589235 | −0.15 | −2. |
| 0.603941 | −0.25 | −3. |
| 0.640706 | −0.25 | −3. |
| 0.154412 | −0.35 | −4. |
| 0.191176 | −0.35 | −4. |
| 0.205882 | −0.45 | −5. |
| 0.242647 | −0.45 | −5. |
| 0.257353 | −0.55 | −6. |
| 0.294118 | −0.55 | −6. |
| 0.308824 | −0.65 | −7. |
| 0.345588 | −0.65 | −7. |
| 0.360294 | −0.75 | −8. |
| 0.397059 | −0.75 | −8. |
| 0.411765 | −0.85 | −9. |
| 0.448529 | −0.85 | −9. |
| 0.463235 | −0.95 | −10. |
| 0.5 | −0.95 | −10. |